%% file: Parameter_estimation_PDE_SCL.tex
\documentclass[final,5p,times,twocolumn]{elsarticle}

\usepackage{amsmath,amssymb,amsthm}
\usepackage{graphicx,epstopdf}
\usepackage{euscript,framed}

\input{header}

\journal{}

\begin{document}

\begin{frontmatter}

\title{Adaptive identification of SISO linear infinite-dimensional systems \tnoteref{funding}}
\tnotetext[funding]{This work was supported by the Science and Engineering Research Board, DST India, via the grant ECR/2017/002583.}
\fntext[]{S. Chattopadhyay ({\tt\small rik.sudipta@gmail.com}), S. Sukumar ({\tt\small srikant@sc.iitb.ac.in}) and V. Natarajan ({\tt\small vivek.natarajan@iitb.ac.in}) are with the Systems and Control Engineering Group, Indian Institute of Technology Bombay, Mumbai 400076, India.}

\author[]{Sudipta Chattopadhyay, Srikant Sukumar and Vivek Natarajan}

\begin{abstract}
We propose an adaptive algorithm for identifying the unknown parameter in a linear exponentially stable single-input single-output infinite-dimensional system. We assume that the transfer function of the infinite-dimensional system can be expressed as a ratio of two infinite series in $s$ (the Laplace variable). We also assume that certain identifiability conditions, which include a persistency of excitation condition, hold. For a fixed integer $n$, we propose an update law driven by real-time input-output data for estimating the first $n+1$ coefficients in the numerator and the denominator of the transfer function. We show that the estimates for the transfer function coefficients generated by the update law are close to the true values at large times provided $n$ is sufficiently large (the estimates converge to the true values as time and $n$ tend to infinity). The unknown parameter can be reconstructed using the transfer function coefficient estimates obtained with $n$ large and the algebraic expressions relating the transfer function coefficients to the unknown parameter. We also provide a numerical scheme for verifying the identifiability conditions and for choosing $n$ sufficiently large so that the value of the reconstructed parameter is close to the true value. The class of systems to which our approach is applicable includes many partial differential equations with constant/spatially-varying coefficients and distributed/boundary input and output. We illustrate the efficacy of our approach using three examples: a delay system with four unknown scalars, a 1D heat equation with two unknown scalars and a 1D wave equation with an unknown spatially-varying coefficient.
\end{abstract}

\begin{keyword}
Adaptive identification \sep finite-dimensional approximation \sep irrational transfer function \sep partial differential equation \sep persistency of excitation.
\end{keyword}

\end{frontmatter}


\section{Introduction}
\label{sec1}
\setcounter{equation}{0} 

Linear infinite-dimensional systems are widely used to model physical plants and identifying the unknown parameter in these models is a problem of fundamental interest in many areas of engineering. There are two distinct approaches to parameter identification: non-adaptive and adaptive. In the non-adaptive approach 
input-output data of the plant is collected over a specific time-interval. A scalar cost function quantifying the difference between the collected output and the output of a plant model which uses a parameter estimate is constructed. Then an iterative minimization algorithm is employed to find the value of the parameter estimate which minimizes the cost function and this minimizer is expected to be the true value of the parameter. In the adaptive approach, a differential equation in time called the parameter update law is driven by the real-time plant input-output data and generates estimates for the unknown model parameter. These estimates are guaranteed to converge to the true value asymptotically in time provided certain persistency of excitation conditions are satisfied \cite{DeRo:94}. While more complex parameter estimation problems can potentially be considered in the non-adaptive framework, the adaptive approach has certain distinct advantages. Firstly, since adaptive algorithms use real-time input-output data they can be used to track a slowly time-varying parameter and also to identify sudden jumps in the parameter value \cite{BaDe:98}, \cite[Chapter 4]{IoSu:96}, \cite{KaRaPiJe:19}. Secondly, unlike non-adaptive algorithms, adaptive algorithms are typically accompanied by persistency of excitation conditions which guarantee global parameter convergence \cite{OrBe:00}. Finally, adaptive parameter estimation algorithms can potentially be used in the design of adaptive controllers \cite{BaScDeRo:97}, \cite{Kug:10}. In this work, we propose a novel adaptive parameter identification algorithm for a class of exponentially stable single-input single-output (SISO) infinite-dimensional systems.  

Early works on the adaptive parameter identification of infinite-dimensional systems, see \cite{BaScDeRo:97}, \cite{DeRo:94a}, \cite{HoBe:94}, \cite{OrBe:00} and references therein, assumed that the full state of the system can be measured. These works introduced and formally established the notion of persistency of excitation for parameter convergence in the context of infinite-dimensional systems. Building on these works \cite{BoKa:16} and \cite{Kug:10} proposed adaptive identification algorithms assuming only partial state measurements, which is also restrictive in the infinite-dimensional setting. Under the more realistic assumption that finite-dimensional outputs are measured, adaptive identification algorithms have been developed for certain types of infinite-dimensional systems in the following papers (and the references therein): \cite{GoOrKo:07} and \cite{NaReXi:14} have considered SISO finite-dimensional systems with unknown delays; 1D hyperbolic partial differential equations (PDEs) with unknown constants in the boundary have been considered in \cite{AnAa:17} and \cite{BiMe:17}; 
positive real infinite-dimensional systems have been considered in \cite{CuDeIt:03} and 1D parabolic PDEs have been considered in \cite[Chapter 12]{SmKr:10} under the structural condition that the measured output multiplies the unknown parameter. 

The recent set of works \cite{KaPiRaUs:19}, \cite{KaPiRaUs:20} and  \cite{KaRaPiJe:19} have proposed adaptive identification algorithms assuming finite-dimensional (often scalar) outputs that are applicable to a larger class of infinite-dimensional systems than those considered so far in the literature. In \cite{KaRaPiJe:19}, a baseline algorithm and a  simplified algorithm with harmonic excitation are presented. Both the algorithms are local in nature and require the initial guess of the unknown parameter to be sufficiently close to its actual value. The adaptation law in both the algorithms contains a gradient term which can be computed only if a closed-form expression of the system transfer function is known. So these algorithms cannot be applied to PDEs with spatially-varying coefficients. In \cite{KaPiRaUs:19} the adaptation law in the baseline algorithm is suitably normalized to obtain finite-time convergence of the parameter estimates and in \cite{KaPiRaUs:20} it is shown that the duration of convergence (finite-time) can be decreased by introducing time-varying adaptation gains.

A natural approach to the design of control/estimation algorithms for an infinite-dimensional system is to approximate it using $n^{\rm th}$-order finite-dimensional systems, then design algorithms for the $n^{\rm th}$-order systems and finally show that the algorithm designed for $n$ sufficiently large can be applied to the infinite-dimensional system to achieve the desired control/estimation objective. This approach has been used in the literature for stabilization \cite{BaKr:02}, \cite{BaKu:84}, \cite{BoBaKr:03}, \cite{Mor:94} and motion-planning \cite{ChNa:2020}, \cite{UtMeKu:2010}. In this work we apply it to estimate the unknown parameter in a class of exponentially stable SISO infinite-dimensional systems. We suppose that the transfer function of the system can be expressed as a ratio of two infinite series in $s$ (the Laplace variable). Considering the $n^{\rm th}$-order transfer function obtained by truncating these series suitably, we propose an update law (a differential equation) for estimating the first $n+1$ coefficients in the numerator and the denominator of the original transfer function. The update law is implemented using real-time measurements of the input and the output of the infinite-dimensional system. 
Under certain assumptions, which include a persistency of excitation condition, we show that as $n$ tends to infinity the asymptotic (in time) estimates of the transfer function coefficients generated by the update law converge to the actual values. Using the estimates obtained by implementing the update law with $n$ sufficiently large and the algebraic expressions relating the transfer function coefficients to the unknown parameter, we then reconstruct the unknown parameter. We remark that typically only the first few transfer function coefficients are used for parameter reconstruction.

The class of infinite-dimensional systems that can be considered using our approach described above includes many linear exponentially stable PDEs with constant/spatially-varying coefficients and distributed/boundary input and output. We remark that when the unknown parameter in the PDE is spatially-varying we can still employ our identification algorithm to obtain estimates for the transfer function coefficients, and these estimates will be accurate provided certain assumptions hold. However, in this case, reconstructing the unknown parameter from the transfer function coefficients will be difficult. The reason for this is that the relationship between the transfer function coefficients and the unknown parameter, while simple in the case of a constant parameter, can be very complicated in the case of a spatially-varying parameter. We illustrate all this using an example in Section \ref{sec5} which addresses the reconstruction of a linearly-varying parameter. The initial condition for the update law in our approach described above can be chosen arbitrarily; this is useful from a practical standpoint. Unlike many earlier works, we also provide a numerical scheme for verifying the persistency assumptions under which our convergence results hold. This scheme also enables us to choose an $n$ such that the estimates for the transfer function coefficients obtained using the update law designed for the chosen $n$ are close to their actual values.

The rest of the paper is organized as follows. Section \ref{sec2} contains the detailed problem statement. We present our identification algorithm in Section \ref{sec3} and a numerical approach in Section \ref{sec4} for verifying the main assumption in Section \ref{sec3}. In Section \ref{sec5} we illustrate our results using three examples: the first example is a second-order system with unknown scalars and unknown output delay, the second example is a 1D heat equation with constant (in space) unknown diffusion and reaction coefficients and boundary input and output and the third example is a 1D wave equation with unknown linearly-varying (in space) elastic rigidity and boundary input and output. We present some concluding remarks and directions for future work in Section \ref{sec6}. \vspace{1mm}


\noindent
{\it Notations:} The Euclidean norm on $\rline^n$ is denoted as $\|\cdot\|_2$. Let $I_n$ be the identity matrix in $\rline^{n\times n}$.
The smallest eigenvalue of a symmetric matrix $A\in\rline^{n\times n}$ is denoted as $\lambda_{\rm min}(A)$. The magnitude and phase of $a\in\cline$ are denoted as $|a|$ and $\angle a$, respectively. For $\gamma \in \rline$, let $\cline^+_\gamma = \{s \in \cline \m|\m {\rm Re}\m s > \gamma\}$ and let $\overline{\cline^+_\gamma}$ be the closure of $\cline^+_\gamma$ in $\cline$. When $\gamma = 0$, we drop the subscript. Let $L^2(a,b)$ be the space of real-valued square-integrable functions on the interval $(a,b)$ and let $H^n(a,b)\subset L^2(a,b)$ denote the usual Sobolev space of order $n$. Let $L^2_{\text{loc}}([0,\infty);\rline^n)$ be the space of $\rline^n$-valued locally square-integrable functions on $[0,\infty)$. For any $\gamma\in\rline$, we take
$L^2_\gamma([0,\infty); \rline^n)$ to be the set
$$ \Big\{ \phi \in L^2_{\text{loc}}([0,\infty);\rline^n) \m\Big|\m \int_{0}^\infty e^{-2\gamma t} \m \|\phi(t)\|_2^2 \m\dd t  < \infty \Big\}$$
with the norm of $\phi\in L^2_\gamma([0,\infty);\rline^n)$ being the square root of the integral in the above expression.  Let $L^\infty([0,\infty);\rline^n)$ be the space of $\rline^n$-valued functions essentially bounded on $[0,\infty)$. The norm of $f\in L^\infty([0,\infty);\rline^n)$ is $\|f\|_{L^\infty}=\sup_{t\in[0,\infty)} \|f(t)\|_2$. The $k^{\rm th}$ time derivative of a $k$-times continuously differentiable function of time $u$ is written as $u^{(k)}$.

\section{Problem Statement} \label{sec2} \setcounter{equation}{0} 

Consider an exponentially stable single-input single-output linear time-invariant system $\Sigma$ whose state evolution is governed by a differential equation on a possibly infinite-dimensional state space.
We make the following natural assumptions about $\Sigma$:
\begin{enumerate}
 \item[(i)] In the absence of input, the output of $\Sigma$ corresponding to any given initial state is in $L^2_\gamma([0,\infty);\rline)$ for some $\gamma<0$.
 \item[(ii)] When the input to $\Sigma$ is $\sin (\omega t)$ for some $\omega\geq0$, its output is the sum of $a\sin (\omega t+\phi)$ for some $a,\phi\in\rline$ and a function in $L^2_\gamma([0,\infty);\rline)$ for some $\gamma<0$.
 \end{enumerate}
Suppose that $\Sigma$ has a transfer function $\GGG$ which is a bounded (in the sup-norm) analytic function on $\cline^+_\gamma$ for some $\gamma<0$. So for any input $u\in L^\infty([0,\infty);\rline)$, the output $y$ of $\Sigma$ corresponding to this input and zero initial condition satisfies \vspace{-1mm}
$$ \hat y(s)=\GGG(s) \hat u(s) \FORALL s\in\cline^+. \vspace{-1mm}$$
Here {\em hat} denotes the Laplace transform. We also suppose that there exist coefficients $p_k, q_k \in \rline$ such that
\begin{equation} \label{eq:TF}
 \GGG(s) = \frac{\sum_{k=0}^\infty p_k s^k}{\sum_{k=0}^\infty q_k s^k} \FORALL s\in\overline{\cline^+}
\end{equation}
and constants $c_0, c>0$ such that
\begin{equation} \label{eq:coeff_est}
 |p_k|+|q_k| \leq c_0\frac{c^k}{k!} \FORALL k\geq0.
\end{equation}
The transfer function of many PDE models, including transport equations and stable 1D heat and wave equations with appropriate input and output, can be expressed in the form of the ratio shown in \eqref{eq:TF} with the coefficients satisfying the estimate in \eqref{eq:coeff_est} for some $c_0, c>0$, see examples in Section \ref{sec5}. Note that \eqref{eq:coeff_est} guarantees the convergence of the series in the numerator and denominator in \eqref{eq:TF} for each $s\in\overline{\cline^+}$.

Let $\Sigma$ contain an unknown (time-invariant) parameter $\Theta$. Then some of the coefficients $p_k$ and $q_k$ of $\GGG$ in \eqref{eq:TF} may depend on $\Theta$ and will therefore be unknown. We address the following identification problem in this paper.

\begin{framed} \vspace{-3mm}
\begin{problem}\label{prob:iden}
Identify the unknown coefficients $p_k$ and $q_k$ in \eqref{eq:TF} using only the input $u$ (chosen appropriately) and the corresponding output $y$ of $\Sigma$ and reconstruct $\Theta$ using the identified coefficients. \vspace{-4mm}
\end{problem}
\end{framed}

We remark that the above problem cannot be solved by letting all the coefficients $p_k$ and $q_k$ to be unknown since if there exists one set of coefficients such that \eqref{eq:TF} and \eqref{eq:coeff_est} hold, then there exist infinitely many such sets of coefficients. However, when some of the coefficients in \eqref{eq:TF} are known, then there may exist a unique set of unknown coefficients such that \eqref{eq:TF} and \eqref{eq:coeff_est} hold and this would render the above problem meaningful (see Remark \ref{rm:uniquerep} for more details).

In Section \ref{sec3} we propose an adaptive identification algorithm for estimating the unknown coefficients $p_k$ and $q_k$ of $\GGG$ under certain assumptions. We demonstrate the efficacy of our algorithm and the reconstruction of the unknown parameter $\Theta$ (from the identified coefficients of $\GGG$) using three infinite-dimensional systems in Section \ref{sec5}.

\section{Identification Algorithm} \setcounter{equation}{0} \label{sec3} 

In this section, we present an approach for identifying the unknown coefficients $p_k$ and $q_k$ of $\GGG$ in \eqref{eq:TF}. In our approach, we fix an integer $n>0$ and drive $\Sigma$ using an appropriate input signal $u=u_n$ (which has certain properties determined by $n$) and denote the corresponding output $y$ of $\Sigma$ by $y_n$. We propose an adaptive identification algorithm, obtained by modifying the one with an integral cost function presented in \cite[Section 4.3.5]{IoSu:96}, for estimating the unknown coefficients multiplying $s^k$ with $k\leq n$ in \eqref{eq:TF}. The algorithm uses only the input signal $u_n$ and the output signal $y_n$. We show in Theorem \ref{th:param_conv} that under certain conditions, which include a persistency of excitation condition, the estimates for the  coefficients of $\GGG$ obtained using our algorithm converge to their actual values as $n\to\infty$.

First we present our identification algorithm. Fix $n\in\nline$ and some $\omega_n>0$. Take the input of $\Sigma$ to be \vspace{-1mm}
\begin{equation} \label{eq:input}
 u_n(t) = \sum_{m=1}^{n+1} \sin(m\omega_n t) \FORALL t\geq0. \vspace{-1mm}
\end{equation}
Let $y_n$ be the output of $\Sigma$ corresponding to this input and zero initial condition, i.e.
\begin{equation} \label{eq:outpuyG}
\hat y_n(s)= \GGG(s) \hat u_n(s) \FORALL s\in\cline^+.
\end{equation}
From the properties of $\Sigma$ it follows that $y_n=y_{n, ss}+y_{n,tr}$, where $y_{n,tr}\in L^2_\gamma([0,\infty);\rline)$ for some $\gamma<0$ and \vspace{-1mm}
$$ y_{n,ss}(t) =  \sum_{m=1}^{n+1} |\GGG(j m\omega_n)| \sin(m\omega_n t+\angle\GGG(j m\omega_n)) \quad \forall t\geq0. \vspace{-1mm}$$
We refer to $y_{n,ss}$ as the steady-state of $y_n$. Let $\Lambda_n$ be an $(n+1)^{\rm th}$-order stable linear system with single input, $(n+1)$ outputs and transfer function matrix given as \vspace{-1mm}
\begin{equation}\label{eq:Fn}
 \FFF_n(s) = ((n+1)\omega_n)^{n+1}\frac{\bbm{ 1 & s &\! \cdots\! & s^n}^\top}{(s+(n+1)\omega_n)^{n+1}}. \vspace{-1mm}
\end{equation}
We define $\Phi_{u_n}$ and $\Phi_{y_n}$ to be the outputs of $\Lambda_n$ when the inputs are $u_n$ and $y_n$, respectively, and the initial condition is zero. So $\hat{\Phi}_{u_n}(s)= \FFF_n(s) \hat {u}_n(s)$ and $\hat{\Phi}_{y_n}(s)= \FFF_n(s) \hat {y}_n(s)$ for all $s\in\cline^+$. Let \vspace{-1mm}
\begin{equation}\label{eq:Phi_n}
\Phi_n(t) = \bbm{\Phi_{u_n}(t)  \\ -\Phi_{y_n}(t)} \FORALL t\geq0. \vspace{-1mm}
\end{equation}
Then naturally \vspace{-1mm}
\begin{equation}\label{eq:hatPhi_n}
\hat \Phi_n(s) = \bbm{\FFF_n(s) \hat u_n(s) \\ -\FFF_n(s) \hat y_n(s)} \FORALL s\in\cline^+. \vspace{-1mm}
\end{equation}
Since $u_n$ and the steady-state of $y_n$ are periodic functions and $\Lambda_n$ is a stable filter it follows that $\Phi_n=\Phi_{n,ss}+ \Phi_{n,tr}$, where  $\Phi_{n,ss}$ is the periodic steady-state of $\Phi_n$ and $\Phi_{n,tr}\in L^2_\gamma([0,\infty);\rline^{2n+2})$ for some $\gamma<0$.

We collect the coefficients in the numerator and denominator of $\GGG$ in \eqref{eq:TF} up to order $n$ in a single vector $\beta_n$: \vspace{-1mm}
\begin{equation}\label{eq:beta_n}
 \beta_n = [p_0 \ \ p_1 \ \cdots \ p_n \ \ q_0 \ \ q_1 \ \cdots \ q_n]^\top. \vspace{-1mm}
\end{equation}
Let the estimated value of the vector $\beta_n$ at time $t$ be \vspace{-1mm}
\begin{equation}\label{eq:est_beta_n}
 \bar{\beta}_n(t) = [\bar{p}_0(t) \ \ \bar{p}_1(t) \ \cdots \  \bar{p}_n(t) \ \ \bar{q}_0(t) \ \ \bar{q}_1(t) \ \cdots \ \bar{q}_n(t)]^\top, \vspace{-1mm}
\end{equation}
where $\bar p_i(t)$ and $\bar q_i(t)$ are estimated values of $p_i$ and $q_i$ at time $t$. If a coefficient $p_i$ is known, then we take $\bar p_i(t)=p_i$ and if $q_i$ is known, then we take $\bar q_i(t)=q_i$. Let $\alpha_n$ be the vector obtained by dropping all the known coefficients from $\beta_n$. So we get
\begin{equation} \label{eq:alpha_n}
  \alpha_n = [p_{a_1} \ \ p_{a_2} \ \cdots \ p_{a_{r_1(n)}} \ \ q_{b_1} \ \ q_{b_2} \ \cdots \ q_{b_{r_2(n)}}]^\top,
\end{equation}
where $0\leq a_1 < a_2 <  \cdots a_{r_1(n)} \leq n$ and $0\leq b_1 < b_2<  \cdots b_{r_2(n)}\leq n$. Let $r(n)=r_1(n)+r_2(n)$. The estimated value of the vector $\alpha_n$ at time $t$ (obtained by dropping entries from $\bar\beta_n(t)$) is
\begin{align}
 \!\!\bar{\alpha}_n(t) &= [\bar p_{a_1}(t) \ \ \bar p_{a_2}(t) \ \cdots \ \bar p_{a_{r_1(n)}}(t)  \nonumber\\
 &\hspace{22mm} \bar q_{b_1}(t) \ \ \bar q_{b_2}(t) \ \cdots \ \bar q_{b_{r_2(n)}}(t)]^\top. \label{eq:est_alpha_n}
\end{align}
In Example 5.1 in Section \ref{sec5} we consider a system for which all the coefficients in the numerator and two of the coefficients in the denominator of \eqref{eq:TF} are unknown. In Examples 5.2 and 5.3 we consider systems for which only the coefficients in the denominator of \eqref{eq:TF} are unknown.

Consider the following cost function: \vspace{-1mm}
\begin{equation}\label{eq:cost}
 J_n(t) = \bar{\beta}_n^\top(t)\m\bigg[ \int_{t-\frac{2\pi}{\omega_n}}^{t} \Phi_n(\tau)\Phi_n^\top(\tau) \dd\tau\bigg]\m \bar\beta_n(t) \vspace{-1mm}
\end{equation}
for all $t\geq0$. In the above expression we take $\Phi_n(t)=0$ for $t<0$.
We update our estimates  $\bar p_i(t)$ and $\bar q_i(t)$ for the unknown coefficients $p_i$ and $q_i$ in $\alpha_n$ via the following update law: \vspace{-1mm}
\begin{equation}\label{eq:update}
 \dot{\bar \alpha}_n(t) = -\Gamma\, \Bigg[ \frac{\partial J_n(t)}{\partial \bar \alpha_n(t)}\Bigg]^\top \FORALL t\geq0, \vspace{-1mm}
\end{equation}
where the scalar $\Gamma>0$ is the adaptation gain.

Next we present the assumption under which we prove our main convergence result in Theorem \ref{th:param_conv}. Recall $p_k$ and $q_k$ from \eqref{eq:TF}, $\Phi_n(t) \in \rline^{2n+2}$ in \eqref{eq:Phi_n} and $r(n)$ defined below \eqref{eq:alpha_n}. Let $\phi_n(t)\in \rline^{r(n)}$ be the vector obtained by dropping the entries of $\Phi_n(t)$ from the same positions from which the entries of $\beta_n$ were dropped to obtain $\alpha_n$ in \eqref{eq:alpha_n}. Then $\phi_n=\phi_{n,ss}+ \phi_{n,tr}$, where $\phi_{n,ss}$ is the $2\pi/\omega_n$-periodic steady-state of $\phi_n$ and $\phi_{n,tr}\in  L^2_\gamma([0,\infty);\rline^{r(n)})$ for some $\gamma<0$.

\begin{framed} \vspace{-3mm}
\begin{assumption}\label{as:PE}
For some known sequence of frequencies $(\omega_n)_{n=1}^\infty$ satisfying $(n+1)\omega_n \geq 1$ for all $n$, there exist sequences of positive scalars $(\kappa_n)_{n=1}^\infty$ and $(T_n)_{n=1}^\infty$ and an integer $N>0$ such that \vspace{-1mm}
\begin{equation} \label{eq:PEcon}
 \int_{t-\frac{2\pi}{\omega_n}}^{t} \phi_{n}(\tau) \phi_{n}^\top(\tau) \dd\tau \geq  \kappa_n I_{r(n)} \quad \forall\, n>N, \ \ \,\forall t>T_n,
\end{equation}
\begin{equation} \label{eq:PEratio}
 \lim_{n\to\infty}\frac{\sum_{k=n+1}^{\infty}(|p_k|+|q_k|) (n+1)^{n+k+\frac{5}{2}} \omega_n^{n+k}}{\omega_n \kappa_n}=0.
\end{equation}
\end{assumption}
\m\vspace{-5mm}
\end{framed}

While \eqref{eq:PEcon} is similar to the usual persistency of excitation condition in the adaptive identification literature,  \eqref{eq:PEratio} imposes constraints on the level of persistency $\kappa_n$ in terms of the coefficients $p_k$ and $q_k$ and the frequency $\omega_n$. Since there is no analytical approach for estimating $\kappa_n$ (this is true even in a finite-dimensional setting), Assumption \ref{as:PE} cannot be verified analytically. However, it can be verified numerically without generating input-output data when the unknown coefficients are present either only in the numerator or only in the denominator of $\GGG$ in \eqref{eq:TF}, see Section \ref{sec4} for details. When the unknown coefficients are present both in the numerator and in the denominator of $\GGG$, then input-output data is required for verifying the assumption. We have illustrated the numerical verification of the above assumption using three examples in Section \ref{sec5}.

\begin{remark} \label{rm:uniquerep}
The no pole-zero cancellation condition guarantees a unique representation for the transfer function of finite-dimensional systems  as a rational function whose denominator is a monic polynomial. This renders the problem of identifying the transfer function coefficients meaningful. However, the same condition does not guarantee a unique representation for the irrational transfer function of infinite-dimensional systems as a ratio shown in \eqref{eq:TF}. For example, the transfer function $e^{-s}$ of a delay system can be expressed as a ratio shown in \eqref{eq:TF} with no common roots between the numerator and the denominator in two different ways: \vspace{-1mm}
$$ \GGG_1(s) =\frac{\sum_{k=0}^\infty \frac{(-s)^k}{k!}}{1}, \qquad
 \GGG_2(s) = \frac{1}{\sum_{k=0}^\infty \frac{s^k}{k!}}. \vspace{-1mm} $$
When an irrational transfer function is written as a ratio shown in \eqref{eq:TF} with some of the coefficients known, then it is possible that the rest of the coefficients are uniquely determined. In this case the problem of identifying the unknown coefficients in \eqref{eq:TF} becomes meaningful. While there is no simple test (like the pole-zero cancellation condition) for verifying the above unique determination, Assumption \ref{as:PE} guarantees it indirectly by permitting the identification of the unknown coefficients, see Theorem \ref{th:param_conv}.
\end{remark}

We now present a simple lemma. Recall the coefficients $p_k$ and $q_k$ from \eqref{eq:TF}.

\begin{framed} \vspace{-3mm}
\begin{lemma}\label{lm:1freq}
For some $\omega\geq0$ let $u_{ss}(t)=\sin(\omega t)$ and $y_{ss}(t)=|\GGG(j\omega)|\sin(\omega t+\angle \GGG(j\omega))$. Then \vspace{-1mm}
\begin{equation}\label{eq:1freq_inf}
 \sum_{k=0}^{\infty}q_k y_{ss}^{{(k)}}(t) =  \sum_{k=0}^{\infty}p_ku_{ss}^{{(k)}}(t) \FORALL t\geq0.\vspace{-4mm}
\end{equation}
\end{lemma}
\end{framed}

\begin{proof}
Let $N$ and $D$ denote the numerator and denominator, respectively, of $\GGG$ in \eqref{eq:TF}. It follows from \eqref{eq:coeff_est} that $N(s)$ and $D(s)$ are well-defined for any $s\in\overline{\cline^+}$ and
\begin{equation} \label{eq:lpmNd}
 D(j\omega)|\GGG(j\omega)|e^{j\angle\GGG(j\omega)} = N(j\omega).
\end{equation}
A simple calculation using the expressions for $u_{ss}$, $y_{ss}$ gives
\begin{align}
 &\!\!\!\!\sum_{k=0}^{\infty}p_ku_{ss}^{{(k)}}(t) =
 {\rm Re}(N(j\omega)) \sin(\omega t)+{\rm Im}(N(j\omega)) \cos(\omega t), \label{eq:uss}\\
 &\!\!\!\!\sum_{k=0}^{\infty}q_ky_{ss}^{{(k)}}(t) =
 \Big[{\rm Re}(D(j\omega)) \sin\big(\omega t + \angle \GGG(j\omega)\big) \nonumber\\
 &\qquad\qquad+{\rm Im}(D(j\omega)) \cos\big(\omega t + \angle \GGG(j\omega)\big)\Big] |\GGG(j\omega)|. \label{eq:yss}
\end{align}
Expanding $\sin(\omega t + \angle \GGG(j\omega))$ and $\cos(\omega t + \angle \GGG(j\omega))$ in \eqref{eq:yss} in terms of $\sin(\omega t)$ and $\cos(\omega t)$ and using \eqref{eq:lpmNd} it is easy to verify that the expression on the right sides of \eqref{eq:uss} and \eqref{eq:yss} are the same, i.e. \eqref{eq:1freq_inf} holds.
\end{proof}

Recall $\Phi_{n,ss}$ introduced below \eqref{eq:hatPhi_n} and $\beta_n$ in \eqref{eq:beta_n}. The last $n+1$ entries of $\Phi_{n,ss}$ (determined by the output $y_n$) are related to the first $n+1$ entries of $\Phi_{n,ss}$ (determined by the input $u_n$) via the coefficients $p_k$ and $q_k$ in \eqref{eq:TF}. In the next proposition we express this relation in the form of a linear regression model in which the coefficients vector $\beta_n$ appears linearly and multiplies $\Phi_{n,ss}$ which is known (this is called the {\em plant parametric equation} in finite-dimensions \cite{IoSu:96}). Since the transfer function $\GGG$ in \eqref{eq:TF} is irrational, the regression model contains a residue $\delta_n$ which, under Assumption \ref{as:PE}, converges to zero as $n\to\infty$. This model is the motivation for considering the cost function $J$ defined in \eqref{eq:cost}. Recall $\phi_{n,ss}$ introduced above Assumption \ref{as:PE} and $\kappa_n$ in Assumption \ref{as:PE}.

\begin{framed} \vspace{-3mm}
\begin{proposition}\label{pr:param_model}
For a sequence of frequencies $(\omega_n)_{n=1}^\infty$ for which Assumption \ref{as:PE} holds, $\Phi_{n,ss}$ (which is determined by the choice of $\omega_n$) and $\beta_n$ satisfy \vspace{-1mm}
\begin{equation}\label{eq:param_exp}
 \beta_n^\top\Phi_{n,ss}(t) + \delta_n(t) = 0 \qquad \forall\, n\in\nline, \ \ \forall\, t \geq 0, \vspace{-1mm}
\end{equation}
with
\begin{equation} \label{eq:dnconv}
 \lim_{n\to\infty}\frac{\|\delta_n\|_{L^\infty} \|\phi_{n,ss}\|_{L^\infty}}{\omega_n\kappa_n}=0.  \vspace{-4mm}
\end{equation}
 \end{proposition}
\end{framed}

\begin{proof}
Recall $u_n$ from \eqref{eq:input} and $y_n$ from the text below \eqref{eq:input}. Let $v_n$ and $z_n$ denote the first component of the vector-valued outputs $\Phi_{u_n}$ and $\Phi_{y_n}$ defined above \eqref{eq:Phi_n}. Then $\hat v_n(s)=\EEE_n(s) \hat u_n(s)$ and $\hat z_n(s)=\EEE_n(s) \hat y_n(s)$ for all $s\in\cline^+$, where
\begin{equation} \label{eq:En}
 \EEE_n(s) = \frac{((n+1)\omega_n)^{n+1}}{(s+(n+1)\omega_n)^{n+1}}.
\end{equation}
Furthermore, $v_n=v_{n,ss}+v_{n,tr}$ and $z_n=z_{n,ss}+z_{n,tr}$, where the periodic steady-states $v_{n,ss}$ and  $z_{n,ss}$ are given
by
\begin{align*}
 v_{n,ss}(t) &= \sum_{m=1}^{n+1} |\EEE_n(jm\omega_n)| \sin(m\omega_n t + \angle \EEE_n(jm\omega_n)),  \\
 z_{n,ss}(t) &= \sum_{m=1}^{n+1} |\EEE_n\GGG(jm\omega_n)| \sin(m\omega_n t + \angle \EEE_n\GGG(jm\omega_n)),
\end{align*}
and $v_{n,tr},z_{n,tr}\in L^2_\gamma([0,\infty);\rline)$ for some $\gamma<0$. From the definitions of $\Phi_n$ and $\Phi_{n,ss}$ it follows easily that
\begin{align}
 \Phi_{n,ss}(t) &= \big[v_{n,ss}(t) \ \ v_{n,ss}^{(1)}(t) \ \cdots \ v_{n,ss}^{(n)}(t) \nonumber \\
 &\qquad -z_{n,ss}(t) \ \ -z_{n,ss}^{(1)}(t) \ \cdots \ -z_{n,ss}^{(n)}(t)\big]^\top. \label{eq:Phinss}
\end{align}
Applying Lemma \ref{lm:1freq} to $v_{n,ss}$ and $z_{n,ss}$ gives  \begin{equation} \label{eq:uyinfss}
 \sum_{k=0}^{\infty}q_kz_{n,ss}^{{(k)}}(t) =  \sum_{k=0}^{\infty}p_kv_{n,ss}^{{(k)}}(t) \FORALL t\geq 0.
\end{equation}
Using $\beta_n$ from \eqref{eq:beta_n} and \eqref{eq:Phinss}, we can rewrite \eqref{eq:uyinfss} as
\begin{equation}\label{eq:lpminftemp}
\beta_n^\top\Phi_{n,ss}(t) + \delta_n(t) =0 \FORALL t \geq 0,
\end{equation}
where
\begin{equation}\label{eq:Deltan}
 \delta_n(t) =   \sum_{k=n+1}^{\infty}p_kv_{n,ss}^{{(k)}}(t)\ - \sum_{k=n+1}^{\infty}q_kz_{n,ss}^{{(k)}}(t).
\end{equation}

Using $|\EEE_n(j\omega)|\leq1$ for all $\omega\in\rline^+$ and the expressions of $v_{n,ss}$ and $z_{n,ss}$ we get after a simple calculation that
\begin{equation} \label{eq:vnzn}
 \|v_{n,ss}^{(k)}\|_{L^\infty} + \|z_{n,ss}^{(k)}\|_{L^\infty} \leq C(n+1)^{k+1}\omega_n^k,
\end{equation}
where $C=1+\sup_{s\in\overline{\cline^+}}|\GGG(s)|$. From this estimate it follows that \vspace{-1mm}
\begin{align}
 \|\delta_n\|_{L^\infty} &\leq \sum_{k=n+1}^{\infty} \|p_kv_{n,ss}^{(k)}\|_{L^\infty} + \|q_kz_{n,ss}^{(k)}\|_{L^\infty} \nonumber\\
 &\leq C \sum_{k=n+1}^{\infty}(|p_k|+|q_k|)(n+1)^{k+1}\omega_n^k. \label{eq:dnbound}
\end{align}
We remark that the inequality in \eqref{eq:coeff_est} guarantees via
the ratio test that the second series in \eqref{eq:dnbound} is convergent. Note that $\Phi_{n,ss}$ contains up to $n^{\rm th}$-order derivatives of $v_{n,ss}$ and $z_{n,ss}$, see \eqref{eq:Phinss}, and recall that $\phi_{n,ss}$ is obtained by dropping some entries of $\Phi_{n,ss}$. Hence using \eqref{eq:vnzn} and $(n+1)\omega_n\geq 1$ from Assumption \ref{as:PE} we get that
\begin{equation} \label{eq:phinbound}
 \|\phi_{n,ss}\|_{L^\infty} \leq \|\Phi_{n,ss}\|_{L^\infty} \leq C\sqrt{2}(n+1)^{n+\frac{3}{2}} \omega_n^n.
\end{equation}
The limit in \eqref{eq:dnconv} now follows directly from the above estimate, \eqref{eq:dnbound} and \eqref{eq:PEratio}.
\end{proof}

Recall the coefficient vector $\alpha_n$ in \eqref{eq:alpha_n} and its estimate $\bar{\alpha}_n(t)$ in \eqref{eq:est_alpha_n}. Note that $\bar{\alpha}_n(t)$ is determined by the choice of $\omega_n$. We now present the main result of this section.

\begin{framed}  \vspace{-3mm}
\begin{theorem} \label{th:param_conv}
Suppose that Assumption \ref{as:PE} holds for a sequence of frequencies $(\omega_n)_{n=1}^\infty$. Then the estimates $\bar\alpha_n(t)$ for $\alpha_n$ obtained via the update law \eqref{eq:update} satisfy
\begin{equation}  \label{eq:param_conv}
 \lim_{n \longrightarrow \infty}\limsup_{t \longrightarrow \infty} \|\bar\alpha_n(t)-\alpha_n\|_2 = 0.  \vspace{-4mm}
\end{equation}
\end{theorem}
\end{framed}

\begin{proof}
From the definitions of $\bar\beta_n$ in \eqref{eq:est_beta_n}, $\bar\alpha_n$ in \eqref{eq:est_alpha_n}, $\Phi_n$ in \eqref{eq:Phi_n} and $\phi_n$ introduced above Assumption \ref{as:PE}, it is easy to see that the constant matrix $L_n=\big[\frac{\partial\bar{\beta}_n} {\partial\bar{\alpha}_n}\big]^\top \in \rline^{r(n)\times (2n+2)}$ satisfies $L_n\Phi_n=\phi_n$. Let $\tilde{\alpha}_n = \bar{\alpha}_n-\alpha_n$. From \eqref{eq:cost} and \eqref{eq:update} we get
$$ \dot{\tilde{\alpha}}_n(t)= -2\Gamma L_n\m \bigg[ \int_{t-\frac{2\pi}{\omega_n}}^t \Phi_n(\tau)\Phi_n^\top(\tau) \dd\tau\bigg]\, \bar\beta_n(t). $$
Let $\bar{\beta}_n = \beta_n + \tilde{\beta}_n$ in the above equation. Then using $L_n\Phi_n=\phi_n$ and $\Phi_n^\top\tilde\beta_n=\phi_n^\top\tilde\alpha_n$ (which is easy to check) we get
\begin{align*}
 \dot{\tilde{\alpha}}_n(t) &= -2\Gamma\m \bigg[ \int_{t-\frac{2\pi}{\omega_n}}^t \phi_n(\tau) \phi_n^\top(\tau) \dd\tau\bigg]\, \tilde \alpha_n(t) \\
 &\hspace{10mm} -2\Gamma\m \bigg[ \int_{t-\frac{2\pi}{\omega_n}}^t \phi_n(\tau) \Phi_n^\top(\tau) \dd\tau\bigg]\, \beta_n.
\end{align*}
Writing $\Phi_n(\tau) = \Phi_{n,ss}(\tau) +\Phi_{n,tr}(\tau)$ in the above equation and defining $\beta_n^\top\Phi_{n,ss}(\tau) =-\delta_n(\tau)$ to simplify the resulting expression we get that
\begin{equation} \label{eq:updt_ana}
 \dot{\tilde\alpha}_n(t) = h_n(t) - 2\Gamma\m\bigg[ \int_{t-\frac{2\pi} {\omega_n}}^{t} \phi_n(\tau)\phi_n^\top(\tau)\dd \tau  \bigg]\, \tilde\alpha_n(t)
\end{equation}
where $h_n = h_n^1-h_n^2$ with \vspace{-1mm}
\begin{align*}
 h_n^1(t) &= 2\Gamma\int_{t-\frac{2\pi}{\omega_n}}^t \delta_n(\tau) \phi_n(\tau) \dd\tau,\\
 h_n^2(t) &= 2\Gamma\m\bigg[\int_{t-\frac{2\pi}{\omega_n}}^t \phi_n(\tau)\Phi_{n,tr}^\top(\tau)\dd\tau\bigg]\,\beta_n. \\[-4.5ex] \nonumber
\end{align*}
Noting that $\phi_n(\tau)=\phi_{n,ss}(\tau)+\phi_{n,tr}(\tau)$ with $\phi_{n,ss} \in L^\infty([0,\infty);\rline^{r(n)})$ and $\phi_{n,tr} \in L^2_\gamma([0,\infty);\rline^{r(n)})$ and $\Phi_{n,tr} \in L^2_\gamma([0,\infty);\rline^{2n+2})$ for some $\gamma<0$, it follows from the above expressions that \vspace{-1mm}
\begin{align}
 \limsup_{t \to \infty}\|h_n^1(t)\|_2 &\leq \frac{4\pi\Gamma}{\omega_n} \|\delta_n\|_{L^\infty}\|\phi_{n,ss}\|_{L^\infty}, \label{eq:hn1lim}\\
 \limsup_{t \to \infty}\|h_n^2(t)\|_2 &= 0. \label{eq:hn2lim} \\[-4.5ex] \nonumber
\end{align}
We remark that since $\phi_n$ and $h_n$ are continuous functions of time, there exists a unique solution to the linear time-varying differential equation \eqref{eq:updt_ana} for all initial states $\tilde \alpha_n(0)$.

Consider the Lyapunov function \vspace{-1mm}
\begin{equation}\label{eq:GNLyap}
 V_n(t) = \frac{1}{2}\tilde\alpha_n^\top(t)\tilde\alpha_n(t) \FORALL t\geq0. \vspace{-1mm}
\end{equation}
Taking the time derivative of $V_n$ and using \eqref{eq:updt_ana} we get \vspace{-1mm}
\begin{align}
 \dot V_n(t)  &= -2\Gamma\tilde \alpha_n^\top(t)\m \bigg[\int_{t-\frac{2\pi}{\omega_n}}^{t}\phi_n(\tau)\phi_n^\top(\tau) \dd\tau \bigg]\, \tilde \alpha_n(t)\nonumber\\
 &\qquad\qquad + \tilde\alpha_n^\top(t) h_n(t). \label{eq:Vdot} \\[-4.5ex] \nonumber
\end{align}
Let $N$ and $T_n$ be as in Assumption \ref{as:PE}. Note that $\|\tilde\alpha_n(t)\|^2_2=2 V_n(t)$. Using this and \eqref{eq:PEcon} to bound the terms on the right-side of \eqref{eq:Vdot}, we get that
\begin{equation}\label{eq:dotvv}
 \dot V_n(t) \leq -4\Gamma\kappa_n V_n(t) + \|h_n(t)\|_2\sqrt{2V_n(t)}
\end{equation}
for all $n>N$ and $t>T_n$. Choose $n>N$, $s_n>T_n$ and let \vspace{-1mm}
$$ d_n=\Bigg(\frac{\sup_{t\in[s_n,\infty)} \|h_n(t)\|_2}{\sqrt{2} \Gamma\kappa_n} \Bigg)^2. \vspace{-1mm}$$
From \eqref{eq:dotvv} it follows that $\dot V_n(t)\leq -2\Gamma\kappa_n V_n(t)$ whenever $V_n(t)\geq d_n$ and $t>s_n$. This immediately implies that $V_n(t)\leq d_n$ for all $t$ sufficiently large. Therefore, using $V_n(t)=\|\tilde \alpha_n(t)\|_2^2/2$, we get
$$ \limsup_{t \to \infty} \|\tilde{\alpha}_n(t)\|_2 \leq \frac{\sup_{t\in[s_n,\infty)} \|h_n(t)\|_2}{\Gamma\kappa_n}. \vspace{-1mm}$$
Taking $\lim_{s_n\to\infty}$ on both sides of the above inequality and using \eqref{eq:hn1lim}-\eqref{eq:hn2lim} it follows that \vspace{-1mm}
\begin{equation} \label{eq:alphan_pen}
 \limsup_{t \to \infty} \|\tilde{\alpha}_n(t)\|_2 \leq \frac{4\pi\|\delta_n\|_{L^\infty} \|\phi_{n,ss}\|_{L^\infty}}{\omega_n\kappa_n}. \vspace{-1mm}
\end{equation}
The limit in \eqref{eq:param_conv} now follows from \eqref{eq:dnconv} and \eqref{eq:alphan_pen}.
\end{proof}


\begin{remark} \label{rm:overparam}
From Theorem \ref{th:param_conv} it follows that if $n$ is sufficiently large, then the error $\tilde\alpha_n(t)$ in the estimates for the coefficients is small for large $t$. (For a discussion on how such an $n$ can be chosen in practice, see Remark \ref{rm:smallratio}.) So given an integer $m>0$ we can obtain good estimates for the unknown coefficients $p_k$ and $q_k$ for $k\in\{0,1,\ldots m\}$ by choosing $n$ sufficiently large. Since $p_k$ and $q_k$ depend on and can often be expressed as a simple function of the unknown parameter $\Theta$, we can reconstruct $\Theta$ using the obtained estimates. The choice of $m$ depends on the dimension of $\Theta$, i.e. the number of unknown scalars being estimated. We have illustrated the reconstruction of an unknown parameter in the case of a delay, 1D heat and 1D wave equations in Section \ref{sec5}.
\vspace{-1mm}
\end{remark}

\section{Numerical verification of Assumption \ref{as:PE}} \setcounter{equation}{0}\label{sec4}

In this section, we present an approach for verifying Assumption \ref{as:PE} numerically without generating input-output data when the unknown coefficients are present either only in the numerator or only in the denominator of $\GGG$ in \eqref{eq:TF}. More specifically, given a sequence of frequencies $(\omega_n)_{n=1}^\infty$, we derive an expression for computing $\kappa_n$ for each $n\in\nline$ such that the inequality in \eqref{eq:PEcon} holds with this $\kappa_n$ for all $t>T_n$ and some $T_n>0$, see Propositions \ref{pr:kappan_num} and \ref{pr:kappan_den}.
(We remark that when the unknown coefficients are present both in the numerator and the denominator of $\GGG$, one must compute $\int_{t-\frac{2\pi}{\omega_n}}^t \phi_n(\tau) \phi_n^\top(\tau)\dd\tau$ using input-output data and then find $\kappa_n$ such that \eqref{eq:PEcon} holds, see Example 5.1.) Using the computed $\kappa_n$ and the known upper bounds for $|p_k|+|q_k|$, see the assumption below, it can be verified whether the ratio inside the limit in \eqref{eq:PEratio} is sufficiently close to zero when $n$ is sufficiently large. When the ratio is close to zero for a certain $n$, the error in the estimates for the coefficients generated by our adaptive identification algorithm which uses that $n$ will also be close to zero at large times, see Remark \ref{rm:smallratio}. Consequently, by running our algorithm with that $n$ we can obtain good estimates for the coefficients $p_k$ and $q_k$ of $\GGG$ in \eqref{eq:TF}. \vspace{-1mm}

\begin{framed} \vspace{-3mm}
\begin{assumption}\label{as:pkqkbounds}
There exist known sequences of scalars $(p_k^u)_{k=0}^\infty$ and $(q_k^u)_{k=0}^\infty$ and constants $c_0$ and $c$ such that \vspace{-1mm}
\begin{equation} \label{eq:pkqk}
 |p_k|\leq p_k^u \leq c_0 \frac{c^k}{k!}, \ \ \ \ \quad |q_k|\leq q_k^u \leq c_0 \frac{c^k}{k!} \ \ \ \ \quad \forall\, k\geq0.
\end{equation}
\end{assumption}
\m\vspace{-6mm}
\end{framed}
\vspace{-3mm}

Since some of the coefficients $p_k$ and $q_k$ can depend on the unknown parameter $\Theta$, to verify the above assumption it is necessary to assume certain bounds on $\Theta$. As illustrated in the examples in the next section, see Figure 5, the bounds for $\Theta$ can often be chosen arbitrarily and do not influence the verification of Assumption \ref{as:PE}. However, choosing tighter bounds leads to more efficient implementation of the identification algorithm, see Section \ref{sec5.1}.

Recall $\GGG$, $u_n$, $\Phi_n$ and $\EEE_n$ from \eqref{eq:TF}, \eqref{eq:input}, \eqref{eq:Phi_n} and \eqref{eq:En}. Using \eqref{eq:outpuyG}, \eqref{eq:Fn} and \eqref{eq:hatPhi_n} we get that $\hat\Phi_n$ and $\hat u_n$ satisfy $\hat{\Phi}_{n}(s) = \HHH_{\Phi_n}(s)\hat{u}_n(s)$ for all $s \in \cline^+$, where $\HHH_{\Phi_n}(s)\in \cline^{2n+2}$ is defined as \vspace{-1mm}
$$\HHH_{\Phi_n}(s) = \EEE_n(s)\Big[[1 \,\ s \,\ s^2 \ \cdots\ s^n] \ \ -\GGG(s)[1\,\ s \,\ s^2 \ \cdots\ s^n]\Big]^\top. \vspace{-1mm}$$
Recall that $\alpha_n\in\rline^{r(n)}$ in \eqref{eq:alpha_n} is obtained by dropping the known coefficients of $\GGG$ from $\beta_n\in\rline^{2n+2}$ in \eqref{eq:beta_n}. Let $\HHH_{n}(s)\in \cline^{r(n)}$ be the vector obtained by dropping the entries of $\HHH_{\Phi_n}(s)$ from the same positions from which the entries of $\beta_n$ were dropped to obtain $\alpha_n$. So when all the coefficients in the denominator of $\GGG$ are known then
\begin{equation}\label{eq:Hn_num}
\HHH_n(s) = \EEE_n(s)\bbm{s^{a_1} & s^{a_2} & s^{a_3} & \cdots & s^{a_{r_1(n)}}}^\top \vspace{-1mm}
\end{equation}
and when all the coefficients in numerator of $\GGG$ are known \vspace{-1mm} then
\begin{equation}\label{eq:Hn_den}
\HHH_n(s) = -\GGG(s)\SSS_n(s), \vspace{-1mm}
\end{equation}
with $ \SSS_n(s) = \EEE_n(s)\bbm{s^{b_1} & s^{b_2} & \cdots & s^{b_{r_2(n)}}}^\top$. Here $r_1(n)$, $r_2(n)$, $a_1, a_2,\ldots a_{r_2(n)}$ and $b_1, b_2,\ldots b_{r_2(n)}$ are as introduced in the definition of $\alpha_n$.

The next proposition presents an expression for computing $\kappa_n$ when all the coefficients in the denominator of $\GGG$ are \vspace{-1mm} known.

\begin{framed} \vspace{-3mm}
\begin{proposition} \label{pr:kappan_num}
Suppose that all the coefficients $q_k$ in the denominator of $\GGG$ in \eqref{eq:TF} are known so that $\HHH_n$ is given by \eqref{eq:Hn_num}.
Then for each $n \in \nline$ and $\omega_n>0$ the inequality in \eqref{eq:PEcon} holds with \vspace{-2mm}
\begin{equation}\label{eq:kappa_num_defn}
\kappa_n = \frac{\pi}{2\omega_n} \lambda_{\min}\bigg(\Re\sum_{m=1}^{n+1} \HHH_n(-jm\omega_n){\HHH_n}^\top(jm\omega_n)\bigg) \vspace{-1mm}
\end{equation}
for all $t>T_n$ and some $T_n>0$. \vspace{-4mm}
\end{proposition}
\end{framed}

\begin{proof}
Note that $\Re\big(\HHH_n(-j\omega){\HHH_n}^\top(j\omega)\big)$ is a real symmetric positive semidefinite matrix for any $\omega>0$. So $\kappa_n$ defined in \eqref{eq:kappa_num_defn} is non-negative. Clearly \eqref{eq:PEcon} holds for all $t\geq0$ if this $\kappa_n$ is zero. So to complete the proof of this proposition, in the discussion below, we suppose that $\kappa_n$ in \eqref{eq:kappa_num_defn} is positive. Recall the following simple trigonometric identities: for $k,m\in\nline$, $\theta_k,\theta_m\in\rline$ and $\omega_n>0$,
\begin{enumerate}
\item[(i)] $\int_0^{\frac{2\pi}{\omega_n}} \sin(k\omega_n t + \theta_k)\sin(m\omega_nt  + \theta_m)\m\dd t$ and $\int_0^{\frac{2\pi}{\omega_n}} \cos(k\omega_nt + \theta_k) \cos(m\omega_nt+\theta_m)\m\dd t$ are equal to $\pi/\omega_n$ if $k=m$ and 0 otherwise,
\item[(ii)] $\int_0^{\frac{2\pi}{\omega_n}} \sin(k\omega_nt+\theta_k)\cos(m\omega_nt+\theta_m)\m\dd t=0$.
\end{enumerate}

Fix $n\in\nline$ and $\omega_n>0$. Since all the coefficients in the denominator of $\GGG$ in \eqref{eq:TF} are known, from the definitions of $\phi_n$ and $\phi_{n,ss}$ presented above Assumption \ref{as:PE} and  \eqref{eq:Phinss} it follows that
\begin{equation}\label{eq:phinss_vnss_d}
 \phi_{n,ss}(t) = [v_{n,ss}^{(a_1)}(t) \ \ v_{n,ss}^{(a_2)}(t) \ \ \cdots \ \ v_{n,ss}^{(a_{r_1(n)})}(t)]^\top,
\end{equation}
where $v_{n,ss}$ is as defined above \eqref{eq:Phinss}. Using the definition of $v_{n,ss}$ and the above trigonometric identities we get via a simple calculation that \vspace{-1mm}
\begin{align}
 &\int_{t-\frac{2\pi}{\omega_n}}^t v_{n,ss}^{(a_i)}(t)v_{n,ss}^{(a_k)}(t)\dd t \nonumber\\
 =&\! \sum_{m=1}^{n+1}\!|\EEE_n(jm\omega_n)|^2  \bigg[ \int_{t-\frac{2\pi}{\omega_n}}^t\!\!\! \frac{\dd^{a_i} \sin(m\omega_nt+\angle\EEE_n(jm\omega_n))}{\dd t^{a_i}} \nonumber\\ &\hspace{20mm}\frac{\dd^{a_k} \sin(m\omega_n t +\angle\EEE_n(jm\omega_n))}{\dd t^{a_k}} \dd t\bigg]. \label{eq:innpdt_vnss} \\[-4.5ex]\nonumber
\end{align}
Observe (using the trigonometric identities) that the term inside the summation on the right side of the above expression is equal to the real part of the $(i,k)$ entry of the matrix $\HHH_n(-jm\omega_n) {\HHH_n}^\top(jm\omega_n)$ multiplied by $\pi/\omega_n$. Hence we can conclude using \eqref{eq:phinss_vnss_d} that \vspace{-1mm}
\begin{equation} \label{eq:omega_n_Snum}
 \int_{t-\frac{2\pi}{\omega_n}}^{t}\phi_{n,ss}(\tau)\phi_{n,ss}^\top(\tau) \dd\tau = \frac{\pi}{\omega_n}\Re\bigg(\sum_{m=1}^{n+1}\HHH_n(-jm\omega_n) {\HHH_n}^\top(jm\omega_n)\bigg).
\end{equation}
Any real symmetric matrix minus its minimum eigenvalue multiplied with the identity matrix is positive semidefinite. Applying this property to the matrix on the right side of the above expression it follows using the definition of $\kappa_n$ in \eqref{eq:kappa_num_defn} that \vspace{-1mm}
\begin{align}\label{eq:phinss_kappa}
&\int_{t-\frac{2\pi}{\omega_n}}^{t}\phi_{n,ss}(\tau)\phi_{n,ss}^\top(\tau) \dd\tau \geq 2\kappa_n I_{r_1(n)}. \vspace{-1mm}
\end{align}
Since $\phi_n=\phi_{n,ss}+ \phi_{n,tr}$, where $\phi_{n,ss}$ is a bounded periodic function and $\phi_{n,tr}\in  L^2_\gamma([0,\infty);\rline^{r_1(n)})$ for a $\gamma<0$ (see discussion above Assumption \ref{as:PE}), it is easy to see that \vspace{-1mm}
$$ \lim_{t\to\infty} \int_{t-\frac{2\pi}{\omega_n}}^{t} \big(\phi_n(\tau) \phi_n^\top(\tau) - \phi_{n,ss}(\tau) \phi_{n,ss}^\top(\tau)\big) \dd\tau =0. \vspace{-1mm}$$
So, since $\kappa_n$ in \eqref{eq:kappa_num_defn} is positive (see the beginning of this proof), there exists $T_n>0$ such that \vspace{-1mm}
$$ \int_{t-\frac{2\pi}{\omega_n}}^{t} \big(\phi_n(\tau) \phi_n^\top(\tau) - \phi_{n,ss}(\tau) \phi_{n,ss}^\top(\tau)\big) \dd\tau \geq -\kappa_n I_{r_1(n)} \vspace{-1mm}$$
for all $t>T_n$. It now follows directly from \eqref{eq:phinss_kappa} and the above inequality that the inequality in \eqref{eq:PEcon} holds
with $\kappa_n$ in \eqref{eq:kappa_num_defn} for each $n\in\nline$ and $t>T_n$.
\end{proof}

The following proposition presents an expression for $\kappa_n$ when all the coefficients in the numerator of $\GGG$ are known. \vspace{-1mm}

\begin{framed} \vspace{-3mm}
\begin{proposition} \label{pr:kappan_den}
Let Assumption \ref{as:pkqkbounds} hold. Suppose that all the coefficients $p_k$ in the numerator of $\GGG$ in \eqref{eq:TF} are known so that $\HHH_n$ is given by \eqref{eq:Hn_den}. For each $\omega>0$, let
$$ \GGG^u(j\omega) = \frac{\sum_{k=0}^\infty p_k (j\omega)^k}{\sum_{k=0}^\infty q^u_{2k} \omega^{2k}+j\sum_{k=0}^\infty q^u_{2k+1} \omega^{2k+1}},$$
where $q_k^u$ are as in Assumption \ref{as:pkqkbounds}. Let
$$ \HHH^u_n(j\omega) = \GGG^u(j\omega)\SSS_n(j\omega) \FORALL \omega>0,$$
where $\SSS_n$ is as defined below \eqref{eq:Hn_den}. Then for each $n \in \nline$ and $\omega_n>0$ the inequality in \eqref{eq:PEcon} holds with
\begin{equation}
 \kappa_n = \frac{\pi}{2\omega_n} \lambda_{\min}\bigg(\Re\sum_{m=1}^{n+1} \HHH_n^u(-jm\omega_n) {\HHH_n^u}^\top(jm\omega_n)\bigg) \label{eq:kappa_den_defn}
\end{equation}
for all $t>T_n$ and some $T_n>0$. \vspace{-4mm}
\end{proposition}
\end{framed}

\begin{proof}
Note that $\Re\big(\HHH_n^u(-j\omega){\HHH_n^u}^\top(j\omega)\big)$ is a real symmetric positive semidefinite matrix for any $\omega>0$. So $\kappa_n$ defined in \eqref{eq:kappa_den_defn} is non-negative. Clearly \eqref{eq:PEcon} holds for all $t\geq0$ if this $\kappa_n$ is zero. So to complete the proof of this proposition, in the discussion below, we suppose that $\kappa_n$ in \eqref{eq:kappa_den_defn} is positive.

Fix $n\in\nline$ and $\omega_n>0$. Since all the coefficients in the numerator of $\GGG$ in \eqref{eq:TF} are known, from the definitions of $\phi_n$ and $\phi_{n,ss}$ presented above Assumption \ref{as:PE} and  \eqref{eq:Phinss} it follows that
$$ \phi_{n,ss}(t) = -[z_{n,ss}^{(b_1)}(t) \ \ z_{n,ss}^{(b_2)}(t) \ \ \cdots \ \ z_{n,ss}^{(b_{r_2(n)})}(t)]^\top, $$
where $z_{n,ss}$ is as defined above \eqref{eq:Phinss}. Mimicking the steps followed to derive \eqref{eq:omega_n_Snum} for the $\phi_{n,ss}$ defined in \eqref{eq:phinss_vnss_d} using the definition of $v_{n,ss}$, we can derive the following expression for the $\phi_{n,ss}$ defined above using the definition of $z_{n,ss}$: \vspace{-2mm}
\begin{equation} \label{eq:phinest}
 \int_{t-\frac{2\pi}{\omega_n}}^{t}\phi_{n,ss}(\tau)\phi_{n,ss}^\top(\tau)  \dd\tau
 =\frac{\pi}{\omega_n} \Re\bigg(\sum_{m=1}^{n+1} \HHH_n(-jm\omega_n)\HHH_n^\top(jm\omega_n)\bigg).
\end{equation}
From the definitions of $\GGG$ in \eqref{eq:TF} and $\GGG^u$ in the
statement of this proposition, it is easy to see that $|\GGG(j\omega)|\geq |\GGG^u(j\omega)|$ for any $\omega>0$. Using this and the definitions of $\HHH_n$ in \eqref{eq:Hn_den} and $\HHH^u_n$ in the
statement of this proposition we get $\Re (\HHH_n(-j\omega) \HHH_n^\top(j\omega)) \geq \Re(\HHH_n^u(-j\omega){\HHH_n^u}^\top(j\omega))$ for any $\omega>0$. It now follows from \eqref{eq:phinest} that \vspace{-2mm}
$$  \int_{t-\frac{2\pi}{\omega_n}}^{t}\phi_{n,ss}(\tau) \phi_{n,ss}^\top(\tau)  \dd\tau  \geq \frac{\pi}{\omega_n} \Re\bigg(\sum_{m=1}^{n+1} \HHH_n^u(-jm\omega_n) {\HHH_n^u}^\top(jm\omega_n)\bigg). \vspace{-2mm}$$
From this inequality, using the property of symmetric matrices presented above \eqref{eq:phinss_kappa} and the definition of $\kappa_n$ in \eqref{eq:kappa_den_defn}, we get \vspace{-2mm}
\begin{equation}\label{eq:phinss_kappa_den}
 \int_{t-\frac{2\pi}{\omega_n}}^{t}\phi_{n,ss}(\tau)\phi_{n,ss}^\top(\tau) \dd\tau \geq 2\kappa_n I_{r_2(n)}. \vspace{-2mm}
\end{equation}
Finally, using the arguments presented below \eqref{eq:phinss_kappa} we can conclude that there exists $T_n > 0$ such that \vspace{-1.5mm}
$$ \int_{t-\frac{2\pi}{\omega_n}}^{t} \big(\phi_n(\tau) \phi_n^\top(\tau) - \phi_{n,ss}(\tau) \phi_{n,ss}^\top(\tau)\big) \dd\tau \geq -\kappa_n I_{r_2(n)} \vspace{-1.5mm}$$
for all $t>T_n$. It now follows from \eqref{eq:phinss_kappa_den} and the above inequality that  \eqref{eq:PEcon} holds with $\kappa_n$ in  \eqref{eq:kappa_den_defn} for all $t>T_n$.
\end{proof}

\begin{remark} \label{rm:smallratio}
For a given $n$, recall that $\tilde\alpha_n(t)$ is the error in the estimates for the coefficients generated by our adaptive identification algorithm at time $t$ using that $n$. Denote the ratio inside the limit in \eqref{eq:PEratio} by $\rho_n$ and denote the same ratio, but with $|p_k|+|q_k|$ replaced with $p_k^u+q_k^u$ from Assumption \ref{as:pkqkbounds}, by $\rho_n^u$. So $\rho_n^u\geq\rho_n\geq0$. The limit $\lim_{n\to\infty}\rho_n=0$ in Assumption \ref{as:PE} cannot be verified analytically and must be verified numerically, see the discussion below the assumption. Naturally, numerical verification entails computing the upper bound $\rho_n^u$ of $\rho_n$ for several values of $n$ and ascertaining that $\rho_n^u$ is small when $n$ is large. This practical approach to the verification of the limit turns out to be sufficient for the purposes of the estimation problem addressed in this paper. Indeed, using \eqref{eq:dnbound} and \eqref{eq:phinbound} in \eqref{eq:alphan_pen} it follows that $\limsup_{t \to \infty} \|\tilde{\alpha}_n(t)\|_2 \leq 4\sqrt{2}\pi C^2\rho_n$, where
$C=1+\sup_{s\in\overline{\cline^+}}|\GGG(s)|$ is independent of $n$. So if $\rho_n$ is sufficiently small for some integer $n>0$, then the coefficients estimation error $\tilde{\alpha}_n(t)$ will also be small at large times for that $n$. Therefore to apply the identification algorithm in Section \ref{sec3} to a given system we first construct a plot of $\rho_n^u$ versus $n$ and then fix an $n$ for which $\rho_n^u$ (and consequently $\rho_n$) is close to zero. We have illustrated this using three examples in Section \ref{sec5}, see Figure 1. \vspace{-2mm}
\end{remark}

\section{Numerical examples}\setcounter{equation}{0}
\label{sec5}

Recall from Section \ref{sec2} the linear system $\Sigma$, its transfer function $\GGG$ in \eqref{eq:TF} and the unknown parameter $\Theta$ in $\Sigma$. In this section, we use the identification algorithm proposed in Sections \ref{sec3} and \ref{sec4} to address Problem \ref{prob:iden} introduced in Section \ref{sec2} for three different systems. For each system, we first verify the hypothesis on $\Sigma$ and whether the ratio inside the limit in \eqref{eq:PEratio} is sufficiently close to zero for $n$ sufficiently large. Then by running our adaptive identification algorithm with a sufficiently large $n$, we estimate the transfer function coefficients $p_k$ and $q_k$ of $\GGG$ and simultaneously reconstruct the unknown parameter $\Theta$ from these estimated coefficients.

The first example considers a second-order system with output delay. The system has four unknown scalars and the transfer function $\GGG$ of the system, written as a ratio shown in \eqref{eq:TF}, has unknown coefficients both in the numerator and the denominator. The second example considers a 1D heat equation with 2 unknown scalars. We illustrate in simulations that our algorithm can be used to reconstruct these unknown scalars even when they are slowly-varying in time. The third example considers a 1D wave equation with an unknown linearly-varying (in space) coefficient. The presented examples address Problem \ref{prob:iden} for the systems that they consider. \vspace{2mm}

\noindent
{\bf Example 5.1.} Let $\Sigma$ be the linear system with the following state space model: for $t\geq0$,  \vspace{-1mm}
\begin{align*}
 \bbm{\dot x_1(t) \\ \dot x_2(t)} &= \bbm{0 & 1 \\ -b & -a} \bbm{x_1(t) \\ x_2(t)} +  \bbm{0 \\ 1}u(t), \\[1ex]
 y(t) &= K x_1(t-\tau). \\[-4ex]
\end{align*}
Here $[x_1(t) \ \ x_2(t) \ \ z(t)] \in \rline\times\rline\times L^2(-\tau,0)$ with $[z(t)](s)=x_1(t-s)$ for each $s\in(0,\tau)$
is the state, $u(t)\in\rline$ is the input, $y(t)\in\rline$ is the output, the output delay $\tau = 0.1$, $a=0.3$, $b=1$ and $K=1.5$. Clearly all the eigenvalues of $\sbm{0 & 1 \\ -b & -a}$ have a negative real part and so  $\Sigma$ is an exponentially stable system which satisfies the assumptions (i) and (ii) presented at the beginning of Section \ref{sec2}. Let the parameters $K, a, b$ and $\tau$ be unknown, i.e. $\Theta=[K \ \ a \ \ b \ \ \tau]$. The transfer function of $\Sigma$ is $\GGG(s)=Ke^{-\tau s}/(s^2+as+b)$ which can be written as \vspace{-1mm}
\begin{equation}\label{eq:TF_trans}
 \GGG(s) = \frac{\sum_{k=0}^{\infty} p_k s^k}{s^2+q_1s+q_0} \quad\textrm{with}\ \ p_k = K\frac{(-\tau)^k}{k!},
\end{equation}
$q_1=a$ and $q_0=b$. We assume the following bounds on the unknown parameters $K, a, b$ and $\tau$ so as to verify Assumption \ref{as:pkqkbounds} (see the discussion below the assumption): $|K|\leq10$, $|a|\leq 5$, $|b|\leq 10$ and $|\tau|\leq 0.2$. Then clearly Assumption \ref{as:pkqkbounds} holds with \vspace{-1mm}
$$ p_k^u=10\frac{0.2^k}{k!} \ \ \quad \forall\, k\geq0, \quad q_0^u = 10,  \quad q_1^u = 5, \quad q_2^u=1, \vspace{-1mm}$$
$q_k^u=0$ for all $k\geq3$ and $c_0=c=10$.\vspace{1mm}

Recall $\kappa_n$ from \eqref{eq:PEcon} and $\rho_n^u$ from Remark \ref{rm:smallratio}. Since unknown coefficients are present both in the numerator and the denominator of $\GGG$ in \eqref{eq:TF_trans}, the results in Section \ref{sec4} cannot be used to verify Assumption \ref{as:PE}. Instead we use input-output data for the verification as follows. We first fix a sequence $(\omega_n)_{n=1}^\infty$ with $\omega_n = 1/(n+1)$. Then for $n\in\nline$, we drive $\Sigma$ by letting its input $u$ to be $u_n$ given in \eqref{eq:input} and denote the corresponding output $y$ by $y_n$. Using the input $u_n$, output $y_n$ and the filter in \eqref{eq:Fn} we obtain $\Phi_n$ defined in \eqref{eq:Phi_n} and from that we get $\phi_n$ (which is defined above Assumption \ref{as:PE}). We then compute  $\frac{1}{2}\lambda_{\min}\big(\int_{t-\frac{2\pi}{\omega_n}}^{t} \phi_{n}(s) \phi_{n}^\top(s) \dd s\big)$ (which converges in time to a limiting value since $\phi_n$ converges to a periodic function of period $2\pi/\omega_n$) and assign its value at a sufficiently large time $t$ to be $\kappa_n$. Finally, we compute $\rho_n^u$ using the values of $\kappa_n$, $\omega_n$ and $p_k^u$ and $q_k^u$ for $k\geq n+1$. Figure 1 shows the plot of $\log\rho_n^u$ versus $n$. From the figure it is evident that $\rho_n^u$ is close to $0$ for $n$ large. \vspace{-5mm}

$$\includegraphics[scale=0.575]{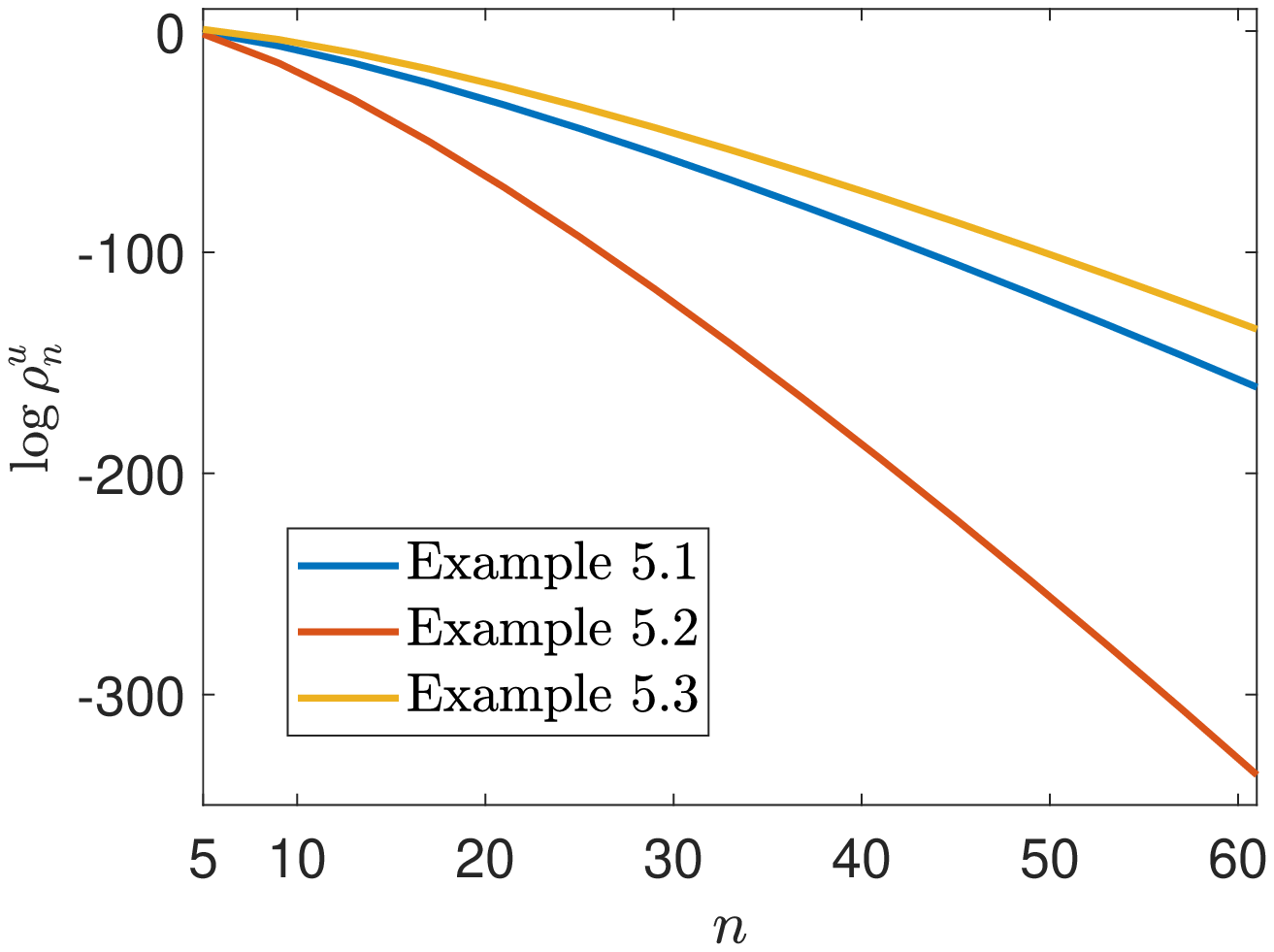} \vspace{-1mm}$$
\centerline{\parbox{3.3in}{
Figure 1. Plot of $\log\rho_n^u$ versus $n$ for Examples 5.1, 5.2 and 5.3. The value of $\rho_n^u$ decreases rapidly as $n$ increases and is close to zero for $n$ large.}}\vspace{-1mm}

$$\includegraphics[scale=0.575]{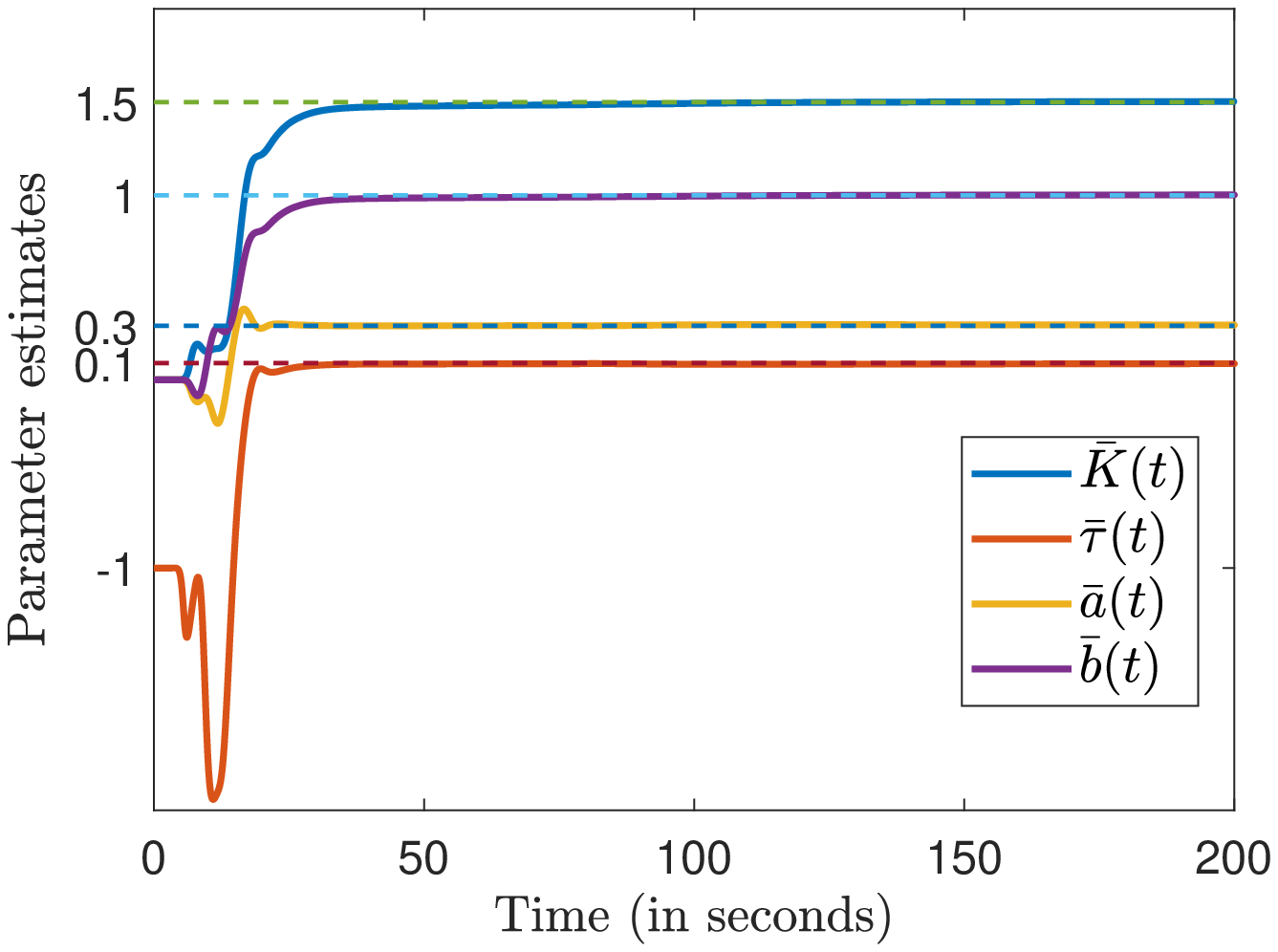} \vspace{-1mm} $$
\centerline{\parbox{3.3in}{
Figure 2. The estimates $\bar K(t)$ for $K$, $\bar\tau(t)$ for $\tau$, $\bar a(t)$ for $a$ and $\bar b(t)$ for $b$ obtained using our identification algorithm approach their actual values $1.5$, $0.1$, $0.3$ and $1$ as $t\to\infty$. At $t=200$ seconds we have $\bar K(t)=1.503$, $\bar\tau(t)=0.097$, $\bar a(t)=0.304$ and $\bar b(t)=1.002$.}}\vspace{3mm}

We implement our identification algorithm in Section \ref{sec3} on $\Sigma$ by taking $n=11$ since $\rho_{11}^u = 3.027\times 10^{-5}$, see Remark \ref{rm:smallratio}. In the update law \eqref{eq:update} we take $\Gamma=50$ and $\bar{\alpha}_{11}(0)$ to be a vector of appropriate dimension with each of its elements being 0.01. Using the estimates for the coefficients $p_0$, $p_1$, $q_0$ and $q_1$ generated by the update law and the expressions \vspace{-1mm}
$$p_0= K, \quad p_1 = -K\tau, \quad q_0 = b, \quad q_1 = a \vspace{-1mm}$$
obtained from \eqref{eq:TF_trans}, we generate estimates for the unknown parameters $K,a,b$ and $\tau$. Figure 2 shows that our estimates for the unknown parameters approach the actual parameter values at large times, as predicted by our theoretical analysis in Theorem \ref{th:param_conv} and Remark \ref{rm:smallratio}. \vspace{2mm}

\noindent
{\bf Example 5.2.} Consider the following 1D heat equation governing the  temperature evolution in a rod of unit length: \vspace{-1mm} for $t\geq0$,
\begin{align}
 &T_t(\xi,t) = \theta\m T_{\xi\xi}(\xi,t) - \lambda\m T(\xi,t) \FORALL \xi \in(0,1), \label{eq:heat1} \\[0.5ex]
 &T_\xi(0,t) = 0, \qquad T_\xi(1,t) = u(t). \label{eq:heat2} \\[-4.5ex] \nonumber
\end{align}
Here $T(\xi,t)$ is the temperature at the location $\xi$ on the rod, $u(t)$ is the input and $\theta,\lambda > 0$ are unknown parameters, i.e. $\Theta=[\theta \ \ \lambda]$. The output \vspace{-1mm} is
\begin{equation}\label{eq:heat3}
  y(t) = T(0,t). \vspace{-1mm}
\end{equation}
The heat equation \eqref{eq:heat1}-\eqref{eq:heat2} along with the output equation \eqref{eq:heat3} can be written as an abstract evolution equation $\Sigma$ on the state space $L^2(0,1)$ as follows: for $t\geq0$, \vspace{-1mm}
\begin{equation} \label{eq:RLSheat}
 \dot x(t) = Ax(t) + Bu(t), \qquad y(t) = C_\Lambda x(t). \vspace{-1mm}
\end{equation}
Here $x(t)\in L^2(0,1)$ is the state. The state operator $A$ is defined as $A \psi = \theta\m\psi_{\xi\xi} - \lambda\m\psi$ for all $\psi \in D(A)$, where \vspace{-1mm}
$$ D(A) = \{\psi \in H^2(0,1) \m\big|\m \psi_\xi(0) = \psi_\xi(1) =0\}. \vspace{-1mm}$$
The control operator $B = \delta_1$, where $\delta_1$ is the dirac pulse at $\xi = 1$. The observation operator $C$ is defined as $C\psi=\psi(0)$ for all $\psi\in D(A)$ and $C_\Lambda$ is the $\Lambda$-extension of $C$. The operator $A$ generates an exponentially stable semigroup on $L^2(0,1)$, the triple $(A,B,C)$ is regular and so $\Sigma$ in \eqref{eq:RLSheat} is a regular linear system. (All this has been discussed in detail in \cite[Section IV]{Na:19} for $\theta=\lambda=1$.)
It follows from the properties of regular linear systems \cite[Section II]{Na:19} that $\Sigma$ is exponentially stable and satisfies assumptions (i) and (ii) presented in the beginning of Section \ref{sec2}. Taking the Laplace transform on both sides of \eqref{eq:heat1}-\eqref{eq:heat3}, the transfer function of $\Sigma$ can be shown to be \vspace{-1.5mm}
$$ \GGG(s) = \frac{1}{\sqrt{\frac{s+\lambda}{\theta}} \sinh\bigg(\sqrt{\frac{s+\lambda}{\theta}}\bigg)} \FORALL s\in\overline{\cline^+}, \vspace{-1.5mm}$$
see \cite[Section IV]{Na:19} for details. Using the Taylor series of $\sinh$ at zero, it follows via a simple calculation that \vspace{-1.5mm}
\begin{equation}\label{eq:heat_TF1}
 \GGG(s) = \frac{1}{\sum_{k=0}^\infty q_ks^k} \vspace{-1.5mm}
\end{equation}
with $q_0 = \sqrt{\frac{\lambda}{\theta}}\sinh \bigg( \sqrt{\frac{\lambda}{\theta}} \bigg)$ and \vspace{-1.5mm}
\begin{equation}\label{eq:heat_TF2}
 q_k= \frac{1}{\theta^{\m k}}\sum_{i=0}^{\infty} \bigg(\frac{\lambda}{\theta}\bigg)^i \binom{k+i}{k}\m\frac{1}{(2k+2i-1)!} \FORALL k\geq1. \vspace{-1.5mm}
\end{equation}
Here $\binom{k+i}{k} = \frac{(k+i)!}{k!\m i!}$. As per the discussion in Section \ref{sec4}, we assume the following bounds on the unknown parameters $\theta$ and $\lambda$ so as to verify Assumption \ref{as:pkqkbounds}: $|\theta|\geq1$ and $|\lambda|\leq 5$. Whenever $\theta$ and $\lambda$ satisfy these bounds, then clearly Assumption \ref{as:pkqkbounds} holds with $p_0^u=1$, $p_k^u=0$ for all $k\geq1$, $q_0^u = \sqrt{5}\sinh\sqrt{5}$, \vspace{-1.5mm}
\begin{equation}\label{eq:heat_qkup}
 q_k^u = \sum_{i=0}^{\infty} 5^i \binom{k+i}{k}\frac{1}{(2k+2i-1)!} \FORALL k\geq1, \vspace{-1.5mm}
\end{equation}
$c_0=e^5=\sum_{i=0}^\infty 5^i\frac{1}{i!}$ and $c=1$. \vspace{1mm}

Next we verify Assumption \ref{as:PE} for $\Sigma$ in \eqref{eq:RLSheat}. Recall $\kappa_n$ from \eqref{eq:PEcon} and $\rho_n^u$ from Remark \ref{rm:smallratio}. Fix a sequence $(\omega_n)_{n=1}^\infty$ with $\omega_n = 1/(n+1)$. Since the unknown coefficients are present only in the denominator of $\GGG$ in \eqref{eq:heat_TF1}, we can use \eqref{eq:kappa_den_defn} in Proposition \ref{pr:kappan_den} along with $q_k^u$ defined earlier to compute $\kappa_n$. Then we can compute $\rho_n^u$ using the values of $\kappa_n$, $\omega_n$ and $p_k^u$ and $q_k^u$ for $k\geq n+1$. Figure 1 shows the plot of $\log\rho_n^u$ versus $n$. From the figure it is evident that $\rho_n^u$ is close to $0$ for $n$ large.

Suppose that the actual value of $\lambda$ is $1.5$ and $\theta$ is as follows: $\theta = 5$ if $t \leq 100$ and $\theta = 6+0.0005t$ if $t > 100$. So $\lambda$ and $\theta$ satisfy the bounds assumed above \eqref{eq:heat_qkup}. We implement our identification algorithm in Section \ref{sec3} on $\Sigma$ by taking $n=9$ since $\rho_{9}^u = 5.624\times 10^{-7}$, see Remark \ref{rm:smallratio}. In the update law \eqref{eq:update} we take $\Gamma=30$ and each entry of the vector $\bar{\alpha}_{9}(0)$ to be $0.1$. Using the estimate for the coefficient $q_0$ generated by the update law and the expression $q_0=\sqrt{\frac{\lambda}{\theta}}\sinh \bigg(\sqrt{\frac{\lambda}{\theta}}\bigg)$ we generate an estimate for $\lambda/\theta$. Then using this estimate, the estimate for $q_1$ from the update law and the formula \vspace{-1.5mm}
\begin{equation*}
q_1 = \frac{1}{\theta} \sum_{i=0}^{\infty}\frac{(1+i)}{(1+2i)!}\m\bigg(\frac{\lambda}{\theta}\bigg)^i \vspace{-1.5mm}
\end{equation*}
from \eqref{eq:heat_TF2} we generate an estimate for $\theta$ first and then for $\lambda$. Figure 3 shows that our estimates for the unknown parameters approach the actual parameter values when they are constant and also track them when they are slowly-varying. \vspace{-2.5mm}


\vspace{-4mm}$$\includegraphics[scale=0.6]{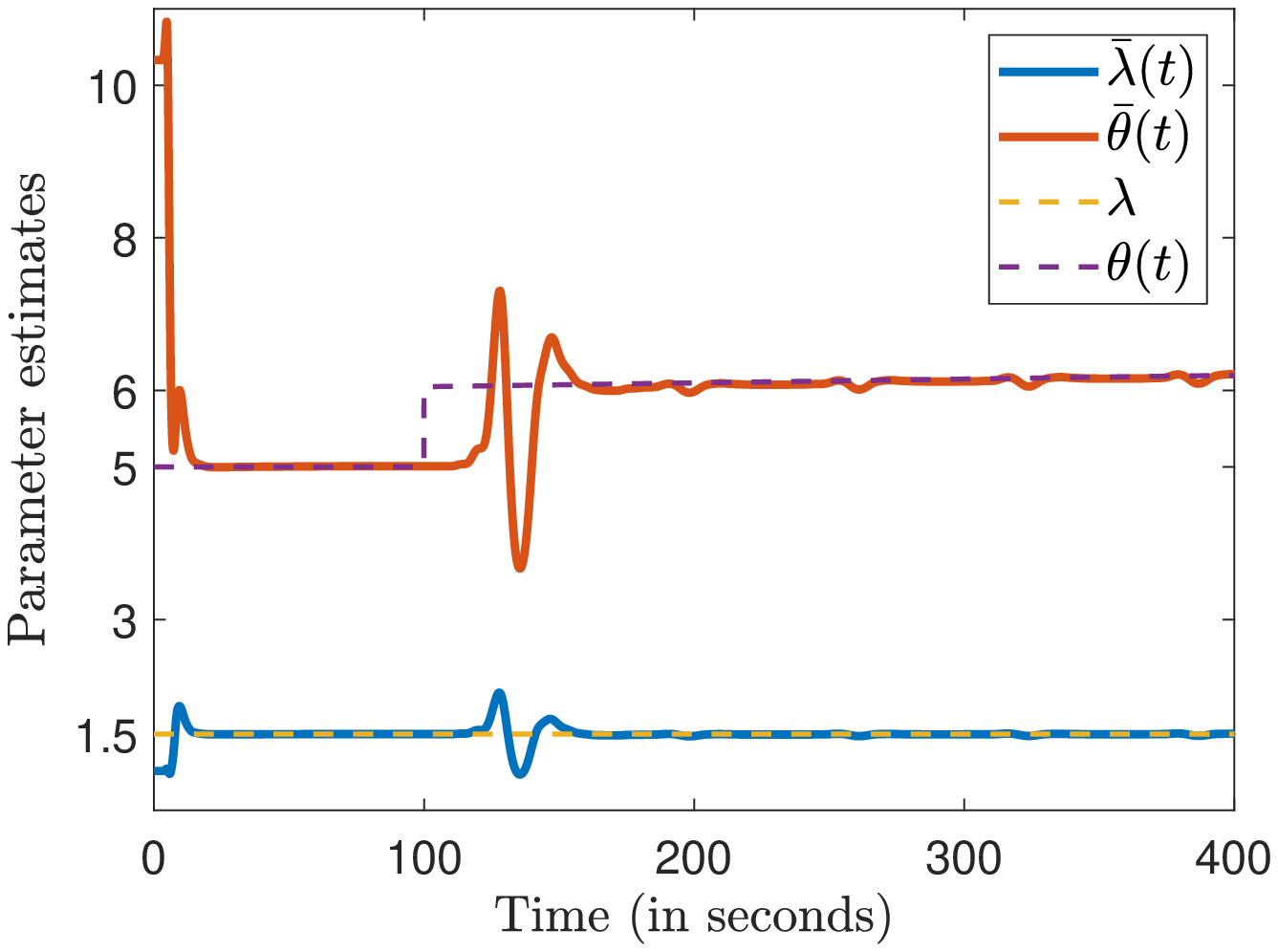} \vspace{-1.5mm}$$
\centerline{\parbox{3.3in}{
Figure 3. The estimates $\bar \theta(t)$ for $\theta(t)$ and $\bar \lambda(t)$ for $\lambda$ obtained using our identification algorithm approach their actual values within 50 seconds. Furthermore, the estimates continue to track the actual values even after $\theta$ undegoes a jump in value at $t=100$ seconds and starts increasing slowly.}}\vspace{4mm}

\noindent
{\bf Example 5.3.} Consider the following 1D wave equation governing the vibration of a string of unit length whose elastic rigidity $EI$ varies linearly in space: for $t \geq 0$, \vspace{-1mm}
\begin{align}
 &w_{tt}(\xi,t) =  (EI(\xi)w_{\xi}(\xi,t))_\xi \FORALL \xi \in(0,1), \label{eq:wave1} \\[0.5ex]
 &w_\xi(0,t) = w_t(0,t), \qquad w(1,t) = u(t). \label{eq:wave2} \\[-4.5ex] \nonumber
\end{align}
Here $w(\xi,t)$ is the displacement of the string at the location $\xi$, $u(t)$ is the input and $EI(\xi)=a+b\xi$ with $a=20$ and $b=10$. The initial condition is $w(\xi,0)=w_t(\xi,0)=0$ for all $\xi\in[0,1]$. The measured output is \vspace{-2mm}
\begin{equation}\label{eq:wave3}
  y(t) = w(0,t). \vspace{-2mm}
\end{equation}
We suppose that the strictly positive linearly-varying function $EI$ or equivalently the scalars $a, b$ are unknown, i.e. $\Theta=[a \ \ b]$.

It is difficult to formulate the wave equation \eqref{eq:wave1}-\eqref{eq:wave2} along with the output equation \eqref{eq:wave3} as an abstract evolution equation. But if the input $u$ is twice differentiable, like the inputs used for identification in this work, then we can transform \eqref{eq:wave1}-\eqref{eq:wave3} to a simpler equation which can be written in the abstract form. So we suppose that $u$ is twice differentiable and define $v(\xi,t) = w(\xi,t)- \xi^2 u(t)$ for all $t \geq 0$ and $\xi \in [0,1]$. It then follows from \eqref{eq:wave1}-\eqref{eq:wave3} that \vspace{-1mm}
\begin{align}
 &v_{tt}(\xi,t) = (EI(\xi)v_{\xi}(\xi,t))_\xi + 2(EI(\xi)\xi)_\xi u(t) - \xi^2 \ddot u(t), \label{eq:wave_vsys1} \\
 &v_\xi(0,t) = v_t(0,t), \qquad v(1,t) = 0, \label{eq:wave_vsys2}\\
 &y(t) = v(0,t) \label{eq:wave_vsys3} \\[-4.4ex] \nonumber
\end{align}
with initial condition $v(\xi,0)=-\xi^2 u(0)$, $v_t(\xi,0)=- \xi^2 \dot u(0)$ for all $\xi\in[0,1]$. Let $H_L^1(0,1)=\{\psi\in H^1(0,1) \m\big|\m \psi(1)= 0\}$ and $X=H_L^1(0,1)\times L^2(0,1)$. We define the norm of $\sbm{\psi_1 \\ \psi_2}\in X$ as \vspace{-1mm}
\begin{equation} \label{eq:normX}
 \Bigg\|\bbm{\psi_1 \\ \psi_2}\Bigg\| = \Bigg(\int_0^1 \Big[EI(\xi)\m\psi_{1,\xi}^2(\xi)+ \psi_2^2(\xi)\Big]\,\dd\xi\Bigg)^{\frac{1}{2}}. \vspace{-1mm}
\end{equation}
Taking $x = \sbm{v \\ v_t}$, \eqref{eq:wave_vsys1}-\eqref{eq:wave_vsys3} can be written as an abstract evolution equation $\Sigma$ on $X$ as follows: for $t\geq0$,
\begin{equation} \label{eq:RLSwave}
 \dot x(t) = Ax(t) + B_1u(t)+B_2\ddot u(t), \qquad  y(t) = Cx(t).
\end{equation}
The state operator $A$ is defined as follows: $A\sbm{v_1 \\ v_2}=\sbm{v_2 \\ (EI v_{1,\xi})_\xi}$ for all $\sbm{v_1 \\ v_2}\in D(A)$, where $D(A)$ is the set \vspace{-1mm}
$$ \Bigg\{\bbm{v_1 \\ v_2} \in H^2(0,1)\times H_L^1(0,1) \m\big|\m v_1(1) = 0, v_{1,\xi}(0)=v_2(0)\Bigg\}. \vspace{-1mm} $$
The control operator $B_1=\sbm{0 \\ 2(EI(\xi)\xi)_\xi}$ and $B_2=\sbm{0 \\-\xi^2}$. The observation operator $C$ is defined as $C\sbm{v_1 \\ v_2}=v_1(0)$ for all $\sbm{v_1 \\ v_2}\in X$. The operator $A$ is Riesz spectral and generates a strongly continuous semigroup \cite{Sh:99} and this semigroup is exponentially stable, see Appendix A for details. Since $B_1$ and $B_2$ are bounded linear operators from $\rline$ to $X$ and $C$ is a bounded linear operator from $X$ to $\rline$ it follows that $(A,[B_1 \ \ B_2],C)$ is a regular triple and so $\Sigma$ in \eqref{eq:RLSwave} (or equivalently \eqref{eq:wave_vsys1}-\eqref{eq:wave_vsys3}) is an exponentially stable regular linear system which satisfies assumptions (i) and (ii) mentioned in the beginning of Section \ref{sec2}.

Taking Laplace transform on both sides of \eqref{eq:wave_vsys1}-\eqref{eq:wave_vsys3} and using the initial conditions given below \eqref{eq:wave_vsys3} we get that for each $s\in\cline^+$, \vspace{-1mm}
\begin{align*}
 & s^2 (\hat v(\xi,s)+\xi^2 \hat u(s)) = \Big(EI(\xi)\m\big(\hat v(\xi,s)+\xi^2\hat u(s)\big)_\xi\Big)_\xi, \\ 
 & \hat v_\xi(0,s) = s\hat v(0,s), \qquad \hat v(1,s) = 0, \\
 &\hat y(s) = \hat v(0,s). \\[-4.4ex]
\end{align*}
Let $\hat{z}_1(\xi,s)=\hat v(\xi,s) + \xi^2 \hat{u}(s)$, $\hat{z}_2(\xi,s) = EI(\xi)(\hat v(\xi,s) + \xi^2 \hat{u}(s))_\xi$,
$\hat z(\xi,s) = \sbm{\hat{z}_1(\xi,s) \\ \hat{z}_2(\xi,s)}$ and $\Ascr(\xi,s) = \sbm{0 & \frac{1}{EI(\xi)} \\ s^2 & 0}$. Then the above equations can be rewritten as \vspace{-1mm}
\begin{align}
 & \hat z_\xi(\xi,s) = \Ascr(\xi,s)\hat z(\xi,s), \label{eq:Peano1}\\
 & \hat{z}_2(0,s) = sEI(0)\hat{z}_1(0,s), \qquad \hat{z}_1(1,s) = \hat u(s), \label{eq:Peano2}\\
 &\hat y(s) = \hat{z}_1(0,s). \label{eq:Peano3} \\[-4.4ex] \nonumber
\end{align}
The solution of \eqref{eq:Peano1}-\eqref{eq:Peano3} is
\begin{equation}\label{eq:Peanosoln}
 \hat z(\xi,s) = \Pscr(\xi,s)\hat z(0,s), \vspace{-1mm}
\end{equation}
where the initial state is $\hat z(0,s)=\hat y(s)\sbm{1 \\ s EI(0)}$ and the state transition matrix $\Pscr(\xi,s)$ is given by the Peano-Baker series \vspace{-2mm}
\begin{align*}
 &\Pscr(\xi,s) = I_2 + \int_0^\xi \Ascr(\tau_1,s)\dd \tau_1 +
 \\  &\hspace{23mm}\int_0^\xi \Ascr(\tau_1,s)\int_0^{\tau_1} \Ascr(\tau_2,s)\dd \tau_2\m\dd \tau_1+\cdots . \\[-4.7ex]
\end{align*}
From \eqref{eq:Peano2} and \eqref{eq:Peanosoln} we get
$\hat u(s)=[1 \ \ 0]\Pscr(1,s)\sbm{1 \\ s EI(0)}\hat y(s)$ which, using the  Peano-Baker series, can be rewritten as $\hat y(s) = \GGG(s)\hat u(s)$ for all $s\in\cline^+$. Here
\begin{equation}\label{eq:wave_TF}
\GGG(s) = \frac{1}{\sum_{k=0}^{\infty}q_k s^k},
\end{equation}
with $q_0 = 1$ and $q_k$ equal to
{\small $$ \int_0^1\!\frac{EI(0)}{EI(\tau_k)}\int_0^{\tau_k}\!\!\int_0^{\tau_{k-1}} \frac{1}{EI(\tau_{k-2})} \cdots \int_0^{\tau_3}\!\!\int_0^{\tau_2} \frac{1}{EI(\tau_1)}\dd \tau_1 \dd \tau_2 \cdots \dd \tau_k $$ }
if $k$ is odd and $q_k$ equal to
{\small $$ \int_0^1 \!\frac{1}{EI(\tau_{k-1})}\int_0^{\tau_{k-1}}\!\! \int_0^{\tau_{k-2}}\!\frac{1}{EI(\tau_{k-3})}\cdots\int_0^{\tau_3}\!\!\int_0^{\tau_2}\! \frac{\tau_1}{EI(\tau_1)}\dd \tau_1 \dd \tau_2 \cdots \dd \tau_{k-1}$$}
\!\!if $k > 0$ is even. We assume the following bounds on the unknown function $EI$ so as to verify Assumption \ref{as:pkqkbounds} (see the discussion below the assumption): $|EI(0)|\leq60$ and $EI(\xi)\geq10$ for all $\xi\in[0,1]$. Then, since in this example $EI(\xi)=20+10\xi$, clearly Assumption \ref{as:pkqkbounds} holds with $p_0^u=1$, $p_k^u=0$ for all $k\geq1$, $q_0^u = 1$, \vspace{-2mm}
\begin{align}
 q_k^u &= \frac{1}{k!} \frac{60}{\sqrt{10}} \bigg(\frac{1}{10}\bigg)^{\frac{k}{2}} \quad \ \text{if} \ k\  \text{is odd}, \label{eq:wave_qk_bound1}\\
 q_k^u &= \frac{1}{k!}\bigg(\frac{1}{10}\bigg)^{\frac{k}{2}}  \quad\quad \text{if} \ k>0\  \text{is even}, \label{eq:wave_qk_bound2}
\end{align}
$c_0=60$ and $c=1$. \vspace{1mm}

Next we verify Assumption \ref{as:PE} for $\Sigma$ in \eqref{eq:RLSwave}. Recall $\kappa_n$ from \eqref{eq:PEcon} and $\rho_n^u$ from Remark \ref{rm:smallratio}. Fix a sequence $(\omega_n)_{n=1}^\infty$ with $\omega_n = 1/(n+1)$. Since the unknown coefficients are present only in the denominator of $\GGG$ in \eqref{eq:wave_TF}, we can use \eqref{eq:kappa_den_defn} in Proposition \ref{pr:kappan_den} along with $q_k^u$ defined above to compute $\kappa_n$. Then we can compute $\rho_n^u$ using the values of $\kappa_n$, $\omega_n$ and $p_k^u$ and $q_k^u$ for $k\geq n+1$. Figure 1 shows the plot of $\log\rho_n^u$ versus $n$. From the figure it is evident that $\rho_n^u$ is close to $0$ for $n$ large.

We implement our identification algorithm in Section \ref{sec3} on $\Sigma$ by taking $n=16$ since $\rho_{16}^u = 3.84\times 10^{-6}$, see Remark \ref{rm:smallratio}. In the update law \eqref{eq:update} we take $\Gamma=50$ and the vector $\bar{\alpha}_{16}(0)$ as follows: first two entries of $\bar{\alpha}_{16}(0)$ are $0.02$ and the rest of the entries are $0$. Computing $q_1$ and $q_2$ using the expressions given below \eqref{eq:wave_TF} and the expression $EI(\xi)= a +b\xi$ we get \vspace{-1mm}
\begin{equation}\label{eq:wave_recon1}
q_1 = \frac{a}{b}\log\bigg(\frac{a+b}{a}\bigg), \qquad q_2 = \frac{1-q_1}{b}. \vspace{-1mm}
\end{equation}
From the estimates for the coefficients $q_1$ and $q_2$ generated by the update law, we first generate an estimate for the unknown parameter $b$ using the second expression in \eqref{eq:wave_recon1} and then for $a$ using the first expression in \eqref{eq:wave_recon1}. Figure 4 shows that our estimates for the unknown parameters approach the actual parameter values at large times as expected from our theoretical analysis. 

\m \vspace{-12mm}
$$\includegraphics[scale=0.575]{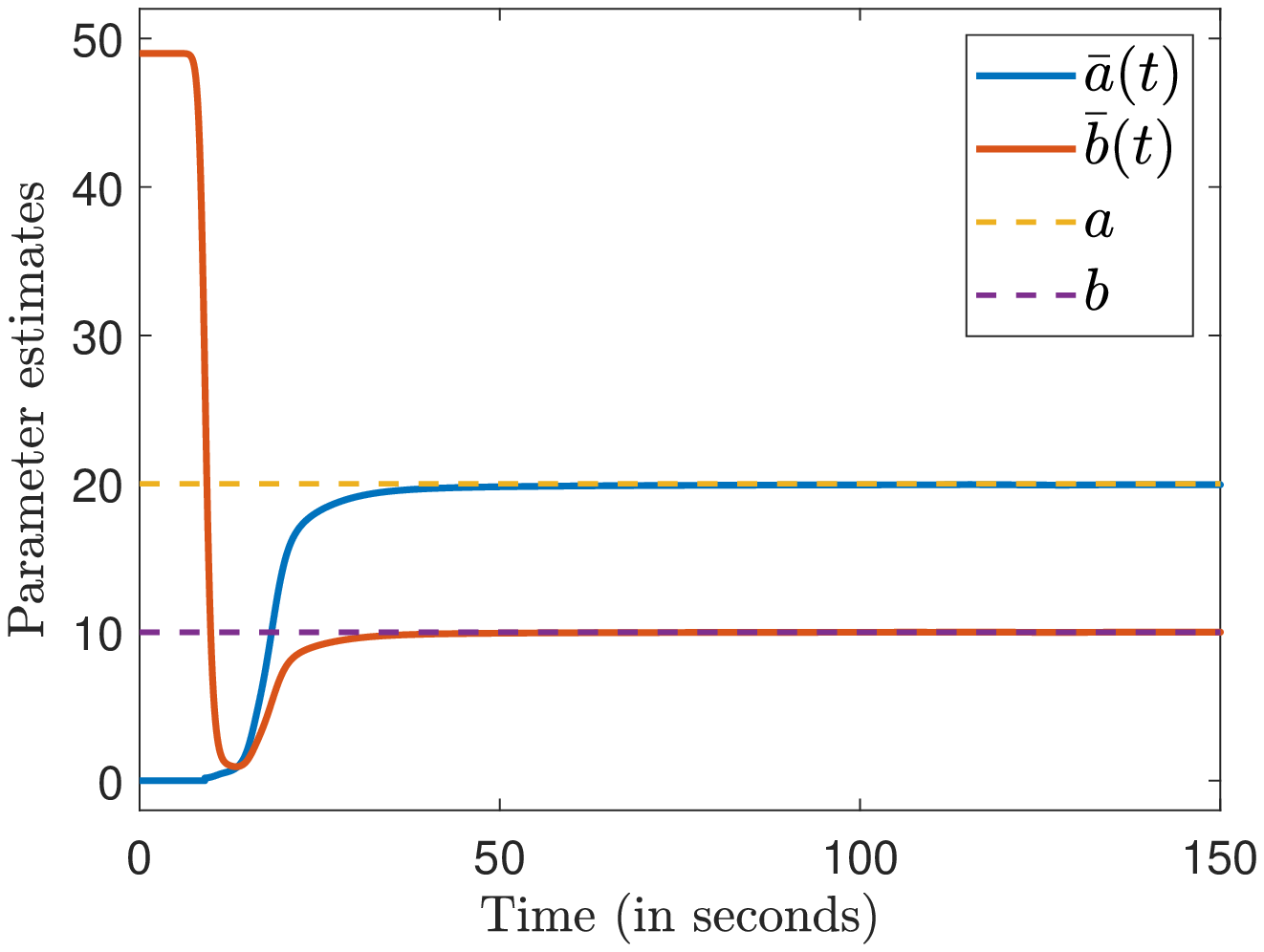} \vspace{1mm}$$
\centerline{\parbox{3.3in}{
Figure 4. The estimates $\bar a(t)$ for $a$ and $\bar b(t)$ for $b$ obtained using our identification algorithm approach their actual values $20$ and $10$ as $t\to\infty$. At $t=150$ seconds we have $\bar a(t)=19.94$ and $\bar b(t)=10.01$.}}

\subsection{Effect of the bounds assumed on $\Theta$} \label{sec5.1}
\vspace{-7mm}

$$\includegraphics[scale=0.575]{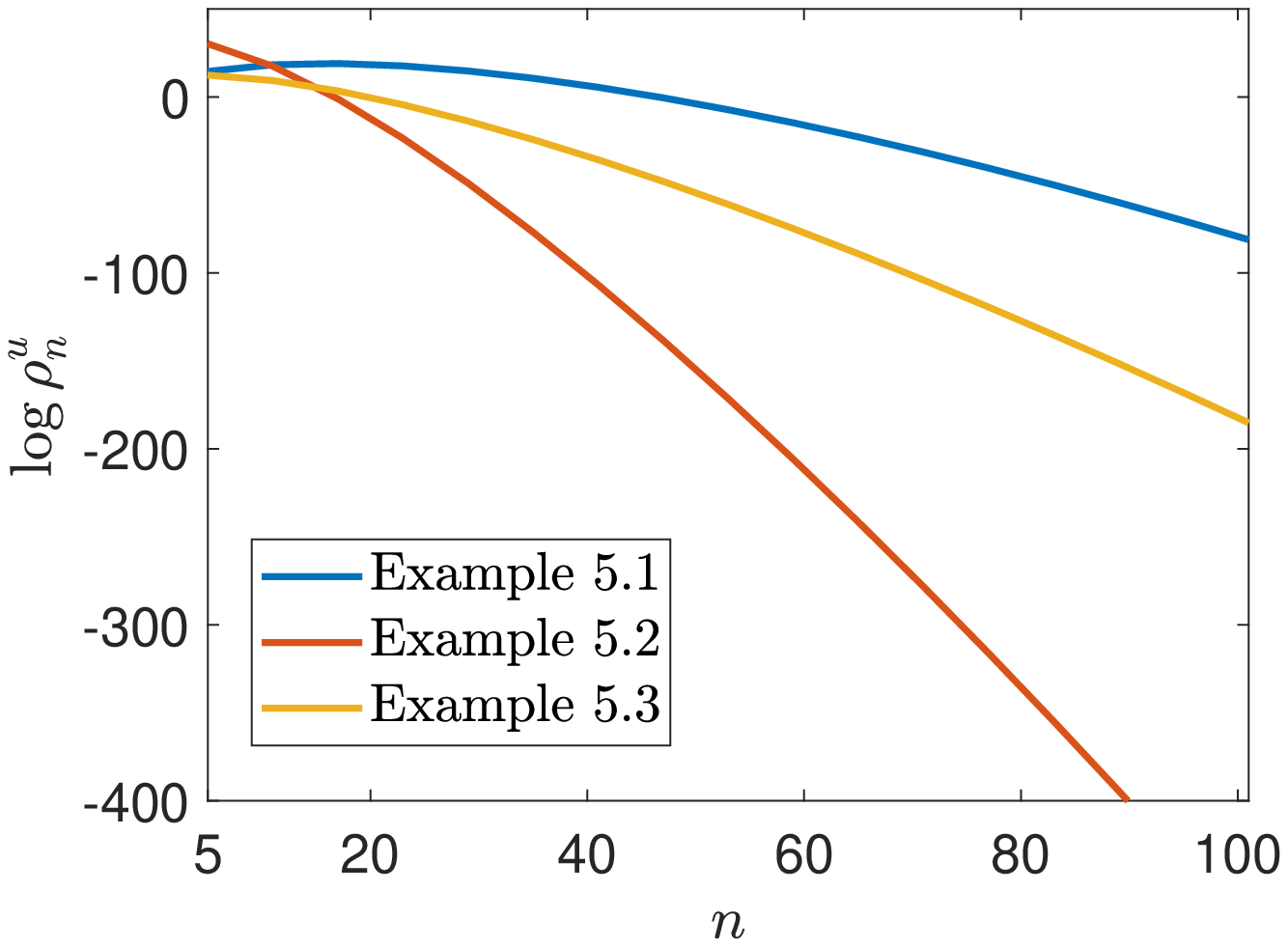} \vspace{-1mm}$$
\centerline{\parbox{3.3in}{
Figure 5. Plot of $\log\rho_n^u$ versus $n$ for Examples 5.1, 5.2 and 5.3 while using less conservative bounds on the unknown parameter $\Theta$. The value of $\rho_n^u$ is close to zero for $n$ large.}}\vspace{3mm}

To verify Assumption \ref{as:pkqkbounds} we have assumed certain arbitrary bounds on the unknown parameter $\Theta$ in Examples 5.1, 5.2 and 5.3. These bounds determine the values of $p_k^u$ and $q_k^u$ in \eqref{eq:pkqk} and through them affect the computation of $\rho_n^u$ (defined in Remark \ref{rm:smallratio}) and thereby influence the numerical verification of Assumption \ref{as:PE}. However, as seen in the examples, changing the bounds on $\Theta$ only changes certain powers in the numerator of $p_k^u$ and $q_k^u$ and leaves the factorials in the denominator unchanged. It is these factorials that enable the verification of Assumption \ref{as:PE}. To illustrate this we consider the following less conservative bounds for $\Theta$ in Examples 5.1, 5.2 and 5.3 and plot the corresponding $\rho_n^u$ in Figure 5: We take $|K|\leq 25$, $|a|\leq 20$, $|b|\leq 25$ and $\tau \leq 2$ in Example 5.1, $\theta \geq 0.2$ and $\lambda \leq 20$ in Example 5.2 and $EI(0) \leq 150$ and $EI(\xi) \geq 2$ for all $\xi\in[0,1]$ in Example 5.3. As seen in Figure 5, similarly to Figure 1, $\rho_n^u$ is close to $0$ for $n$ large. So choosing less conservative bounds for $\Theta$ does not invalidate the verification of Assumption \ref{as:PE}. However, it leads to larger values of $p_k^u$ and $q_k^u$ and therefore of $\rho_n^u$. So an $n$ selected for implementing our identification algorithm (such that $\rho_n^u$ is less than a certain value) using Figure 5 will be larger than an $n$ selected using Figure 1.
Thus assuming tighter bounds on $\Theta$ leads to a more efficient and less complex (i.e. smaller $n$) implementation of our algorithm.

\section{Conclusions} \label{sec6}

In this paper we have proposed an adaptive algorithm for identifying the transfer function coefficients of exponentially stable single-input single-output infinite-dimensional systems. Our algorithm is implemented using real-time measurements of the input and the output, and these measurements can be either distributed or boundary-valued. Using three examples, we have demonstrated that an unknown (and possibly spatially-varying) parameter in certain 1D PDEs can be reconstructed from the transfer function coefficients identified using our algorithm. In the examples, we have considered constant and linearly-varying parameters. While our identification algorithm itself does not preclude the consideration of a more complex spatially-varying parameter, the challenge lies in the reconstruction of the parameter from the identified transfer function coefficients. Preliminary studies indicate that increasing the number of outputs and utilizing optimization algorithms at the reconstruction stage are two promising approaches for addressing this challenge. Both these approaches will benefit from an increase in the speed of convergence of the identification algorithm. We plan to explore these ideas in a future work. \vspace{-1mm}

\appendix
\section{} \vspace{-1mm}
Here we briefly sketch a proof of the exponential stability of the state operator $A$ introduced below \eqref{eq:RLSwave} in Example 5.3.
Our proof mimics those presented in \cite[Section 5]{NaZhWeFr:19}.

Recall that $EI(\xi)=20+10\xi$. Suppose that the input $u=0$. Then since $A$ generates a semigroup on $X$, it follows that for any given initial state $x(0)=\sbm{v(\cdot,0)\\v_t(\cdot,0)}\in X$, there exists a unique state trajectory $x(t)=\sbm{v(\cdot,t)\\v_t(\cdot,t)}\in X$ for \eqref{eq:wave_vsys1}-\eqref{eq:wave_vsys2} starting from $x(0)$. Along the state trajectory define $V(t) = H(t)+\eta(t)$ with \vspace{-2mm}
\begin{align*}
 H(t) &= \frac{1}{2}\int_0^1 \Big(EI(\xi)\m v_\xi^2(\xi,t) + v_t^2(\xi,t) \Big) \dd\xi, \\
 \eta(t) &= \frac{1}{2} \int_0^1 (\xi - 1)v_\xi(\xi,t)v_t(\xi,t)\dd\xi. \\[-4.6ex]
\end{align*}
Clearly using Young's inequality we have $0.5 H(t) \leq V(t)\leq 2 H(t)$ for all $t\geq0$. Differentiating $V$ with respect to time and using \eqref{eq:wave_vsys1}-\eqref{eq:wave_vsys2} (with $u=0$) it follows after a simple calculation via integration by parts that \vspace{-1mm}
$$ \dot V(t) \leq -0.5 H(t)-14.75 v_t^2(0,t) \leq -0.25 V(t), \vspace{-1mm} $$
which implies that \vspace{-1mm}
\begin{equation}\label{eq:wave_V_est}
 V(t) \leq V(0)e^{-0.25 t} \FORALL t \geq 0. \vspace{-1mm}
\end{equation}
Strictly speaking, we can differentiate $V$ only along differentiable state trajectories of \eqref{eq:wave_vsys1}-\eqref{eq:wave_vsys2}. But we can show using regularity results (such as Theorem 1.5 in Chapter 6 of \cite{Pa:83}) that the estimate in \eqref{eq:wave_V_est} is valid along all state trajectories of \eqref{eq:wave_vsys1}-\eqref{eq:wave_vsys2} (when $u=0$). It now follows using \eqref{eq:wave_V_est} and the expression $0.5 H(t) \leq V(t)\leq 2 H(t)$ that $H(t) \leq 4 H(0)e^{-0.25t}$. So from \eqref{eq:normX} we get that along state trajectories $\|x(t)\| \leq 2\|x(0)\| e^{-0.125 t}$ for all $t\geq0$, i.e. $A$ generates an exponentially stable semigroup. \vspace{-2mm}


\end{document}

%% file: header.tex
\newtheorem{theorem}{Theorem}[section]

\newtheorem{proposition}[theorem]{Proposition}
\newtheorem{lemma}[theorem]{Lemma}
\newtheorem{assumption}[theorem]{Assumption}

\newtheorem{remark}[theorem]{Remark}

\newtheorem{problem}[theorem]{Problem}
\numberwithin{equation}{section}

\renewcommand{\cline}{{\mathbb C}}

\newcommand{\nline}  {{\mathbb N}}
\newcommand{\rline}  {{\mathbb R}}

\newcommand{\EEE}  {{\bf E}}
\newcommand{\FFF}  {{\bf F}}
\newcommand{\GGG}  {{\bf G}}
\newcommand{\HHH}  {{\bf H}}

\newcommand{\SSS}  {{\bf S}}
\newcommand{\dd}   {{\rm d}\hbox{\hskip 0.5pt}}

\newcommand{\Ascr} {{\cal A}}

\newcommand{\Pscr} {{\cal P}}

\newcommand{\mm}    {{\hbox{\hskip 0.5pt}}}
\newcommand{\m}     {{\hbox{\hskip 1pt}}}

\newcommand{\bluff} {{\hbox{\raise 15pt \hbox{\mm}}}}
\newcommand{\sbluff}{{\hbox{\raise  7pt \hbox{\mm}}}}

\renewcommand{\l}    {{\lambda}}

\newcommand{\FORALL} {{\hbox{$\hskip 11mm \forall \;$}}}

\renewcommand{\Re}   {{\rm Re\,}}

\newcommand{\bbm}[1]{\left[\begin{matrix} #1 \end{matrix}\right]}
\newcommand{\sbm}[1]{\left[\begin{smallmatrix} #1
   \end{smallmatrix}\right]}

%% file: Parameter_estimation_PDE_SCL.bbl
\begin{thebibliography}{00}


\bibitem{AnAa:17} H. Anfinsen and O. M. Aamo, ``Adaptive stabilization of $2\times 2$ linear hyperbolic systems with an unknown boundary parameter from collocated sensing and control,'' {\it IEEE Trans. Autom. Control}, vol. 62, pp. 6237-6249, 2017.

\bibitem{BaKr:02} A. Balogh and M. Krstic, ``Infinite dimensional backstepping-style feedback transformations for a heat equation with an arbitrary level of instability,'' {\it European J. Control}, vol. 8, pp. 165-175, 2002.

\bibitem{BaDe:98} H. T. Banks and M. A. Demetriou, ``Adaptive parameter estimation of hyperbolic distributed parameter systems: non-symmetric damping and slowly time varying systems,'' {\it ESAIM: Control, Optim. and Calculus of Variations}, vol. 3, pp. 133-162, 1998.

\bibitem{BaKu:84} H. T. Banks and K. Kunisch, ``The linear regulator  problem for parabolic systems,'' {\it SIAM J. Control and Optim.}, vol. 22, pp. 684-698, 1984.

\bibitem{BaScDeRo:97} J. Baumeister, W. Scondo, M. A. Demetriou and I. G. Rosen, ``On-Line parameter estimation for infinite-dimensional dynamical systems,'' {\it SIAM J. Control and Optim.}, vol. 35, pp. 678-713, 1997.


\bibitem{BiMe:17} M. Bin and F. D. Meglio, ``Boundary estimation of boundary parameters for linear hyperbolic PDEs,'' {\it IEEE Trans. Autom. Control}, vol. 62, pp. 3890-3904, 2017.

\bibitem{BoKa:16} R. Boiger and B. Kaltenbacher, ``An online parameter identification method for time dependent partial differential equations,'' {\it Inverse Problems}, vol. 32, pp. 1-28, 2016.

\bibitem{BoBaKr:03} D. M. Bo\v{s}kovi\'{c}, A. Balogh and M. Krstic, ``Backstepping in infinite dimension for a class of parabolic distributed parameter systems,'' {\it Math. Control Signals Systems}, vol. 16, pp. 44-75, 2003.

\bibitem{ChNa:2020} S. Chatterjee and V. Natarajan, ``Steady-state to steady-state transfer of PDEs using semi-discretization and flatness,'' {\it Proc. $59^{\rm th}$ IEEE Conf. Decision and Control}, pp. 4454-4459, 2020, Jeju Island, Republic of Korea.

\bibitem{CuDeIt:03} R. F. Curtain, M. A. Demetriou and K. Ito, ``Adaptive compensators for perturbed positive real infinite-dimensional systems,'' {\it Int. J. Appl. Math. Comput. Sci.}, vol. 13, pp. 441-452, 2003.


\bibitem{DeRo:94} M. A. Demetriou and I. G. Rosen, ``On the persistence of excitation in the adaptive estimation of distributed parameter systems,'' {\it IEEE Trans. Autom. Control}, vol. 39, pp. 1117-1123, 1994.

\bibitem{DeRo:94a} M. A. Demetriou and I. G. Rosen, ``Adaptive identification of second-order distributed parameter systems,'' {\it Inverse Problems}, vol. 10, pp. 261-294, 1994.


\bibitem{GoOrKo:07} O. Gomez, Y. Orlov and I. V. Kolmanovsky, ``On-line identification of SISO linear time-invariant delay systems from output measurements,'' {\it Automatica}, vol. 43, pp. 2060-2069, 2007.

\bibitem{HoBe:94} K.-S. Hong and J. Bentsman, ``Application of averaging method for integro-differential equations to model reference adaptive control of parabolic systems,'' {\it Automatica}, vol. 30, pp. 1415–1419, 1994.

\bibitem{IoSu:96} P. A. Ioannou and J. Sun, {\it Robust Adaptive Control}, Dover Publications, Inc., Mineola, 2012.



\bibitem{KaPiRaUs:19} M. N. Kapetina, A. Pisano, M. R. Rapaić and E. Usai, ``Adaptive parameter estimation for infinite-dimensional LTI systems with finite-time convergence,'' {\it Proc. $58^{\rm th}$ IEEE Conf. Decision and Control}, pp. 1722-1727, 2019, Nice, France.

\bibitem{KaPiRaUs:20} M. N. Kapetina, A. Pisano, M. R. Rapaić and E. Usai, ``Adaptive unit-vector law with time-varying gain for finite-time parameter estimation in LTI systems,'' {\it Applied Numerical Math.}, vol. 155, pp. 16-28, 2020.

\bibitem{KaRaPiJe:19} M. N. Kapetina, M. R. Rapaić, A. Pisano and Z. D. Jeličić, ``Adaptive parameter estimation in LTI systems,'' {\it IEEE Trans. Autom. Control}, vol. 64, pp. 4188-4195, 2019.

\bibitem{Mor:94} K. A. Morris, ``Design of finite-dimensional controllers for infinite dimensional systems by approximation,'' {\it J. Math. Systems, Estim. Control}, vol. 4, pp. 1-30, 1994.

\bibitem{Kug:10} P. K\"ugler, ``Online parameter identification without Ricatti-type equations in a class of time-dependent partial differential equations: an extended state approach with potential to partial observations,'' {\it Inverse Problems}, vol. 26, pp. 1-23, 2010.

\bibitem{NaReXi:14} J. Na, X. Ren and Y. Xia, ``Adaptive parameter identification of linear SISO systems with unknown time-delay,''
    {\it Systems $\&$ Control Letters}, vol. 66, pp. 43-50, 2014.

\bibitem{Na:19} V. Natarajan, ``Stabilization of PDE-ODE cascade systems using Sylvester equations,'' {\it Proc. $58^{\rm th}$ IEEE Conf. Decision and Control}, pp. 5906-5911, 2019, Nice, France.

\bibitem{NaZhWeFr:19} V. Natarajan, H.-C. Zhou, G. Weiss and E. Fridman, `` Exact controllability of a class of nonlinear distributed parameter systems using back-and-forth iterations,'' {\it Int. J. Control}, vol. 92, pp. 145-162, 2019.

\bibitem{OrBe:00}  Y. Orlov and J. Bentsman, ``Adaptive distributed parameter systems identification with enforceable identifiability conditions and reduced-order spatial differentiation,'' {\it IEEE Trans. Autom. Control}, vol. 45, pp. 203-216, 2000

\bibitem{Pa:83} A. Pazy, {\it Semigroups of Linear Operators and Applications to Partial Differential Equations}, Springer-Verlag, New York, 1983.


\bibitem{Sh:99} M. A. Shubov, ``The Reisz basis property of the system of root vectors for the equation of a nonhomogeneous damped string: transformation operators method,'' {\it Methods and Appl. Analysis}, vol. 6, pp. 571-592, 1999.


\bibitem{SmKr:10} A. Smyshlyaev and M. Krstic, {\it Adaptive Control of Parabolic PDEs}, Princeton University Press, Princeton, 2010.



\bibitem{UtMeKu:2010} T. Utz, T. Meurer and  A. Kugi, ``Trajectory planning for quasilinear parabolic distributed parameter systems based on finite-difference semi-discretisations,'' {\it Int. J. Control}, vol. 83, pp. 1093-1106, 2010.

\end{thebibliography}
